\documentclass[final]{article}
\usepackage[final]{graphicx}
\usepackage{graphicx}
\usepackage{amssymb}
\usepackage{rotating}
\usepackage{pdflscape}
\usepackage[english]{babel}
\usepackage{color}
\usepackage{bigintcalc}
\def\beq{\begin{eqnarray}}
\def\eeq{\end{eqnarray}}
\usepackage{amsfonts}

\usepackage{seqsplit}

\newtheorem{rule-of-thumb}[theorem]{Definition} 

\begin{document}


\title{Thomson problem in one dimension: minimal energy configurations of $N$ charges on a curve}

\author{Paolo Amore \\
Facultad de Ciencias, CUICBAS, Universidad de Colima,\\
Bernal D\'{i}az del Castillo 340, Colima, Colima, Mexico  \\
paolo.amore@gmail.com 
\and
Martin Jacobo \\
Facultad de Ciencias, Universidad Aut\'onoma de San Luis Potos\'i,  \\
Av. Salvador Nava S/N Zona Universitaria, CP 78290, \\
San Luis Potos\'i, SLP, México \\
martiin.jacobo@gmail.com }

\maketitle

\begin{abstract}
We have studied the configurations of minimal energy of $N$ charges on a curve on the plane,
interacting with a repulsive potential $V_{ij} = 1/r_{ij}^s$, with $s \geq 1$ and $i,j=1,\dots, N$.
Among the examples considered are ellipses of different eccentricity, a straight wire and  a
cardioid. We have found that, for some geometries, multiple minima are present, as well as points of 
unstable equilibrium. For the case of the cardioid, we observe that the presence of the cusp has a dramatic
effect on the distribution of the charges, in the limit $N \gg 1$.
\end{abstract}

\section{Introduction}
\label{sec:intro}

In this paper we study the question of how $N$ equal charges, repelling each other with a potential
$V(r_{ij}) = 1/r_{ij}^s$ ($r_{ij}$ is the euclidean distance between the two charges) and $s \geq 1$, 
distribute themselves on a plane curve (either open or close).
This problem is a variant of the well known Thomson problem~\cite{Thomson1904}, that amounts to finding 
the configurations of $N$ charges on a sphere for which the total electrostatic energy is minimal. 
Although the original motivation of Thomson, to provide a model of the atom, has soon become obsolete with the advent 
of Quantum Mechanics, the interest towards the model has increased over the time, in particular related to the study of topological 
defects \cite{Moore97,Bowick02,Bausch03,Backofen10,Irvine10,Wales13,Azadi14,Stefan16,Jimenez16}, of colloidal systems 
\cite{Bausch03,Lipowsky05}, of the structure of viruses~\cite{Zandi04,Longuet09}
and of the properties of crystals on curved surfaces~\cite{Irvine10}, among others.

Apart these physical applications, the Thomson problem has been used as a benchmark problem in testing global 
optimization algorithms~\cite{Erber91,Glasser92,Morris96, Altschuler97,Altschuler05,Wales09,Bondarenko16,Birtea17}, being a challenging computational problem.
The difficulty in finding the numerical solutions to this problem is due to the rapid growth 
of the number of configurations corresponding to local minima of the electrostatic energy, which  
grows exponentially with $N$, as observed by Calef et al.~\cite{Calef15} (these authors report that,
for $N \approx 200$, there are more than $10000$ stable configurations). 
Examples of global minima obtained with a numerical approach can be found, for instance, in 
Refs.~\cite{Erber91, Glasser92, Altschuler97, Wales09} report (Ref.~\cite{Wales09}, in particular, 
reaches configurations of up to $4352$ charges). 
Moreover, this problem has also been studied in relation to the problem of discretizing manifolds
using minimum energy points~\cite{Saff97,Saff98,Saff04}.

Another variant of Thomson problem considers the optimal arrangement of equal charges inside a disk or under 
the action of a confining potential. Also in this case the computational complexity of the problem is greatly
increased by the exponential growth of the number of local minima, as the number of particles trapped inside
the domain grows.  The properties of such systems have been studied  by different authors, as 
for instance refs.~\cite{Berezin85,Wille85,Peeters94, Peeters95, Nurmela98, Oymak00,Oymak01,Worley06, Moore07}.

The focus of the present paper is on one--dimensional systems, represented by $N$ charges arranged on a plane curve, 
in a way that the total electrostatic energy is globally minimal. Although asymptotic ($N \rightarrow \infty$)  estimates
for the behavior  of the total energy of these systems have been derived~\cite{Martinez04,Borodachov12}, the study 
for finite $N$ has been essentially restricted to the case of a straight needle~\cite{Griffiths96,Griffiths01b}, and
mainly focussed on the continuum limit, $N \rightarrow \infty$. Jackson in particular has studied the charge density of 
a thin straight wire, as the transversal dimension shrinks, and proved that it approaches (slowly) a uniform distribution~\cite{Jackson00,Jackson02} (quite interestingly, in ref.~\cite{Jackson02} Jackson points out that the problem had been already
addressed by Maxwell in one of his last papers \cite{Maxwell}).

We have obtained precise numerical results for different curves (circle, ellipses, straight needle and cardiod) and for
configurations with different number of charges, interacting via a potential $V(r_{ij}) = 1/r_{ij}^s$ and $s \geq 1$.
We have observed the appearance of multiple local minima; for the case of ellipses of different eccentricity we
observe the tendency towards a uniform distribution of charges as $N \rightarrow \infty$, while, for the case
of the cardioid, the cusp affects greatly the charge distribution, particularly in the region of the cusp. 

The paper is organized as follows: in Section \ref{circle} we discuss the case of a unit circle, which is exactly solvable,
and reproduce the correct leading asymptotic behavior of the energy with great accuracy; in Section \ref{results} we present the numerical results for ellipses of different eccentricity, a unit straight needle and a cardioid; finally in Section
\ref{concl} we present our conclusions.

\section{An exactly solvable example: the unit circle}
\label{circle}

Due to the particular symmetry of the problem, a configuration of $N$ charge evenly distributed on the circle corresponds
to a stable configuration of equilibrium. It is easy to convince oneself that, when the charges are distributed 
in such a way that the distance of each charge from its closest neighbors to the left and to the right is constant, the resultant
of all forces acting on that particular charge must be normal to the circle.

We parametrize the curve as
\begin{eqnarray}
{\bf r}(\theta) = \cos\theta \ \hat{i} + \sin\theta \ \hat{j} \hspace{1cm} ; \hspace{1cm} 0 \leq \theta < 2\pi
\end{eqnarray}
and select a suitable configuration corresponding to the angles
\begin{equation}
\theta_k(N) = \frac{2 \pi k}{N} \hspace{1cm} ; \hspace{1cm} k=1, \dots, N
\label{min_circle}
\end{equation}

The distance between two charges in this case is given by
\begin{eqnarray}
r_{ij} = |{\bf r}_i -{\bf r}_j| = 2 \left| \sin \frac{\pi (i-j)}{N}\right|
\end{eqnarray}
and the total energy of the system takes the form
\begin{eqnarray}
E_s = \sum_{i=2}^N \sum_{j=1}^{i-1} \frac{1}{r_{ij}^s} = \frac{N}{2^{s}} \sum_{k=1}^{N-1} \frac{1}{\sin^s \frac{\pi k}{N}} 
\label{en_circle}
\end{eqnarray}
with $s \geq 1$ ($s=1$ is the Coulomb potential).

Since this expression can be evaluated efficiently with high accuracy and limited effort, we may use it 
to estimate the asymptotic expression for $\mathcal{E}_s(N)$ as $N \rightarrow \infty$ and contrast the numerical results
obtained using eq.~(\ref{en_circle}) with the exact asymptotic results of \cite{Martinez04,Borodachov12} (see for example eqns
(4.1)-(4.6) of \cite{Borodachov12}).

In particular, we consider the electrostatic potential, corresponding to $s=1$ and calculate the energies corresponding 
to the configurations for different values of $N$. We have found out that the most accurate fits correspond  to using the form
\begin{eqnarray}2
\left. \frac{\mathcal{E}_1(N)}{N^2 \log N} \right|_{N \gg 1} \approx c_0 + \sum_{k=1}^{k_{max}} \frac{c_k}{N^{2(k-1)} \log N} 
\end{eqnarray}

Using a set of numerical results from $N=370$ to $N=400$ (obtained with $40$ digit accuracy) and using the fit above with 
$k_{max} = 6$, we have obtained the estimates
\begin{eqnarray}
c_0 &=& \underline{0.3183098861837906715377675267} \nonumber \\
c_1 &=& \underline{0.0399902130750533171061531}70 \nonumber  \\
c_2 &=& - 0.0436332312998582394231 \nonumber  
\end{eqnarray}

These results should be compared with the exact results of Refs.~\cite{Martinez04,Borodachov12} ($\Gamma$ is the length of the curve)
\begin{eqnarray}
c_0^{(exact)} &=& \frac{2}{|\Gamma|} = \frac{1}{\pi} \approx 0.31830988618379067153776752674 \nonumber \\
c_1^{(exact)} &=& \Phi_1(\Gamma) + \frac{2}{|\Gamma|} (\gamma - \log 2) \nonumber \\
&=& \frac{\gamma }{\pi}+\frac{\log (2)}{\pi}-\frac{\log (\pi )}{\pi } \approx 0.0399902130750533171061531699456
\end{eqnarray}
where, for the circle
\begin{eqnarray}
\Phi_1(\Gamma) &=& \frac{1}{|\Gamma|^2} \int_0^{2\pi} du \int_0^{2\pi} dt
\left[ \frac{1}{2 \sqrt{\sin ^2\left(\frac{t}{2}-\frac{u}{2}\right)}}-\frac{1}{\min \left(2 \pi
   -\sqrt{(t-u)^2},\sqrt{(t-u)^2}\right)}  \right] \nonumber \\
&=& \frac{2 \log (2)}{\pi }-\frac{\log (\pi )}{\pi }\approx  0.0768923606293969297433251339198
\end{eqnarray}

The two leading coefficients predicted by the formulas of refs.~\cite{Martinez04, Borodachov12} 
are thus reproduced very accurately by the fits of the numerical data (underlined digits in the expressions of $c_0$ and
$c_1$ obtained from the fit agree with the exact results of refs.~\cite{Martinez04, Borodachov12}).

Note that, for even values of $s$, the energy of eq.~(\ref{en_circle}) can be calculated exactly:
\begin{eqnarray}
\mathcal{E}_2(N) &=& \frac{ N \left(N^2-1\right) }{12} \nonumber \\
\mathcal{E}_4(N) &=& \frac{ N \left(N^2-1\right) }{720}  \left(N^2+11\right) \nonumber \\
\dots &=& \dots \nonumber
\end{eqnarray}

\section{Numerical results}
\label{results}

In this section we report and discuss the equilibrium configurations for $N$ charges confined on curves in the plane and
subject to repulsive interactions $V_{ij}=1/r_{ij}^s$, where $r_{ij}$ is the euclidean distance between any two charges and 
$s \geq 1$. The examples that we have considered are ellipses of different eccentricity,
a straight segment of unit length (needle) and a cardioid.

In all cases the equilibrium configurations have been found numerically, with high accuracy, 
applying the Newton method, for different number of charges.

\subsection{Ellipse}
\label{sub:ellipse}

The second example that we consider, is the ellipse described by the parametric equation
\begin{eqnarray}
{\bf r}(\theta) = (b \cos \theta , \sin \theta )  \hspace{1cm} ; \hspace{1cm} 0 \leq \theta < 2 \pi
\end{eqnarray}
where $b>1$ is the major semiaxis. Clearly this domain is symmetric with respect to reflections
about the principal axes ($(x,y) \rightarrow (x,-y)$ and $(x,y) \rightarrow (-x,y)$) and inversion
through the origin ($(x,y) \rightarrow (-x,-y)$).

In this case, the equilibrium configurations can only be found numerically. As for the other examples that we consider 
in this work, Newton's method has been used to find the minima of the electrostatic energy, for a wide range of
number of charges. Typically in all our numerical results the value of the energy has been determined with an accuracy of
$100$ digits.

\begin{figure}[t]
\begin{center}
\includegraphics[width=6cm]{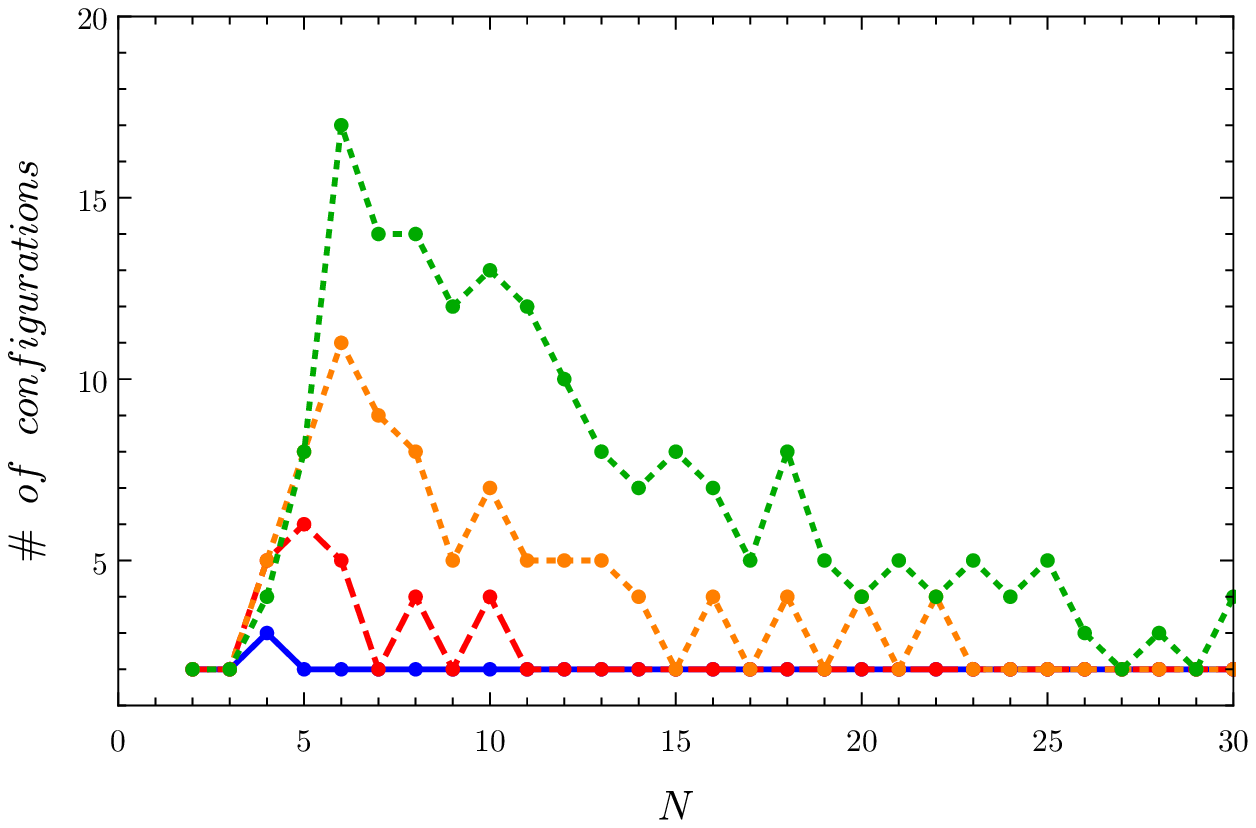} \hspace{1cm} 
\includegraphics[width=6cm]{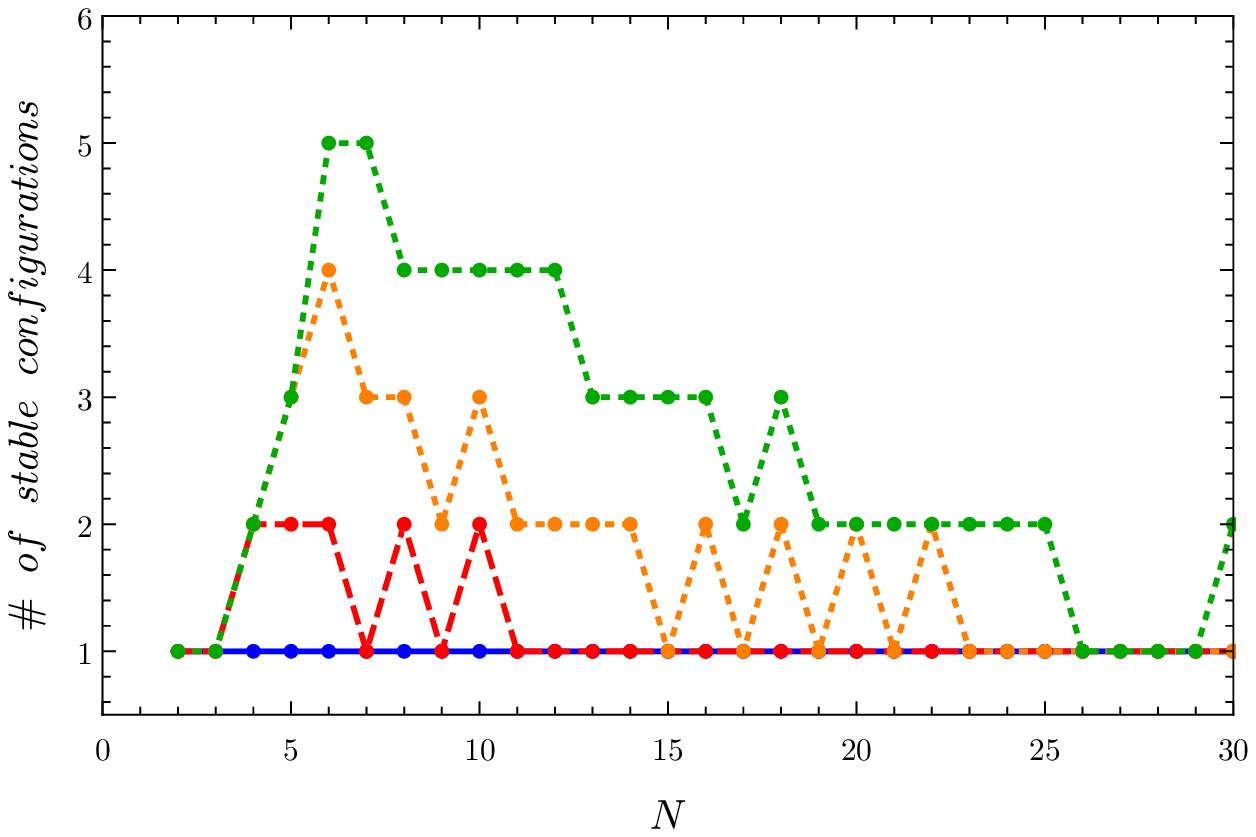} 
\caption{Left plot: Total number of equilibrium configurations on ellipses with major semiaxis $b=2,3,4,5$ (in order of increasing values); 
Right plot: number of stable configurations}
\label{fig_ellipse_1}
\end{center}
\end{figure}

A first interesting discovery that we have made studying the Thomson problem on ellipses of different eccentricity, is that 
the number of equilibrium configurations (both stable and unstable) increases as $b$ takes larger and larger values, 
for a given number of particles on the ellipse. This behavior is displayed in Fig.~\ref{fig_ellipse_1}, where we have plotted the total number of equilibrium configurations (left plot) and the number of stable equilibrium configurations (right plot), as a function of 
the number of charges on the ellipse, repelling each other with a Coulomb force ($s=1$), up to $30$ charges.
One can notice that, as $b$ grows, the number of available configurations also grows, with a peak that  moves to the right.
The number of configurations eventually decreases as $N$ gets sufficiently large, contrary to the behavior observed for 
the Thomson problem on a sphere, where the number of local minima grows exponentially with $N$, for large
$N$~\cite{Calef15}.

\begin{figure}[!htb]
\begin{center}
\includegraphics[width=5cm]{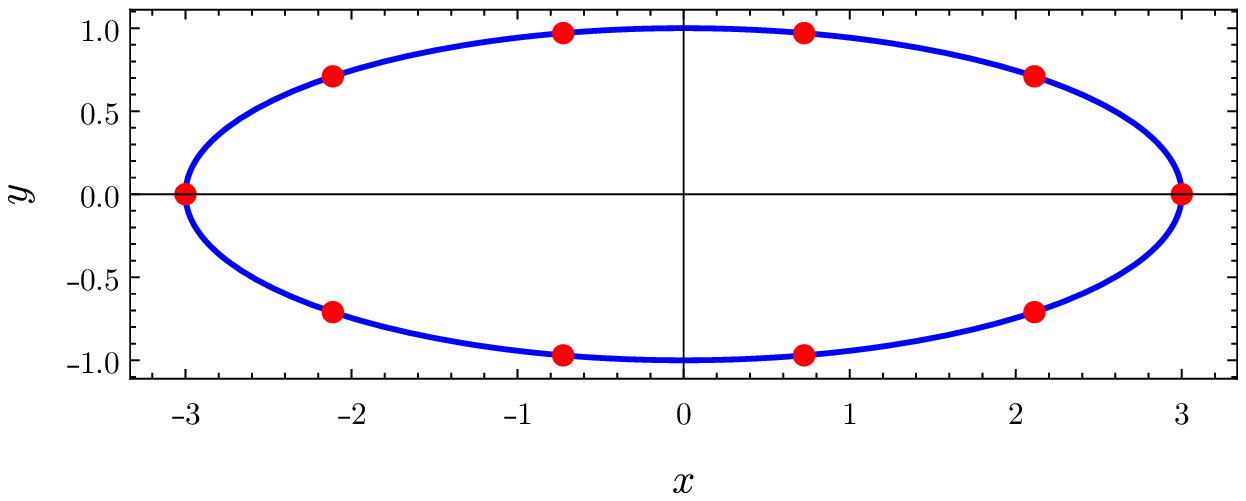} \hspace{1cm} 
\includegraphics[width=5cm]{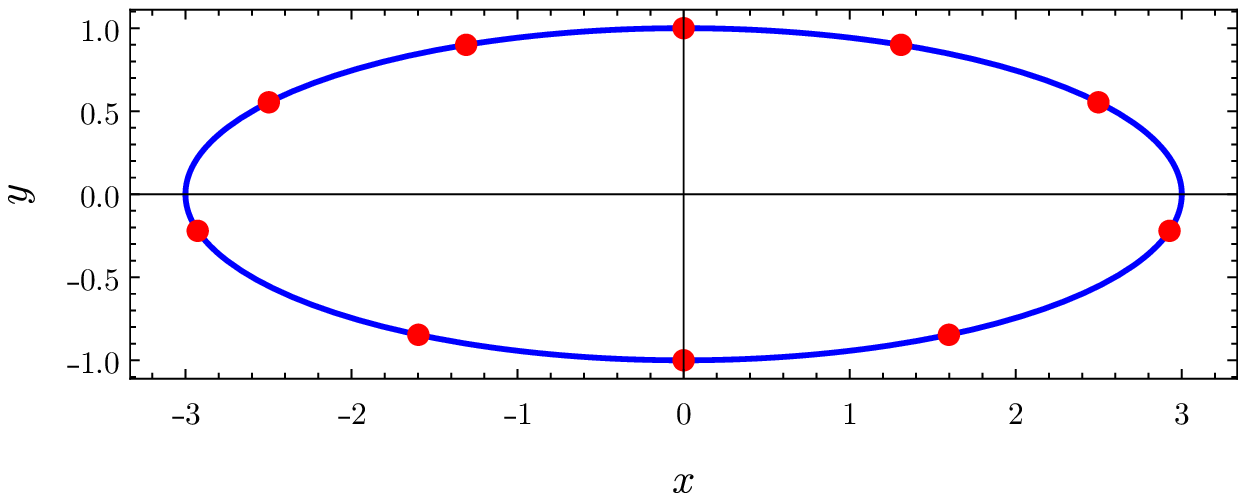} 
\vspace{.5cm}
\\
\includegraphics[width=5cm]{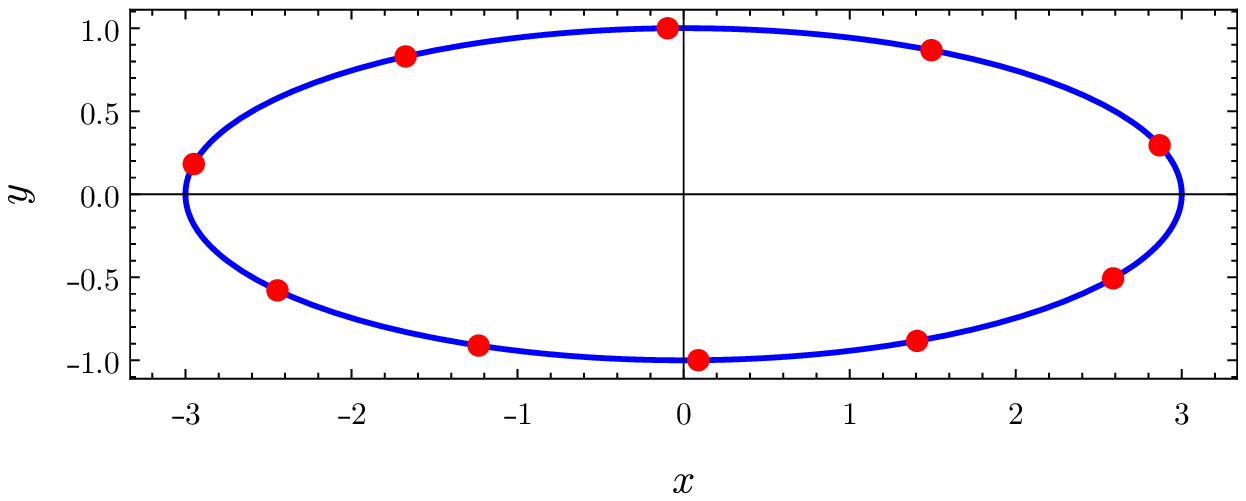} \hspace{1cm} 
\includegraphics[width=5cm]{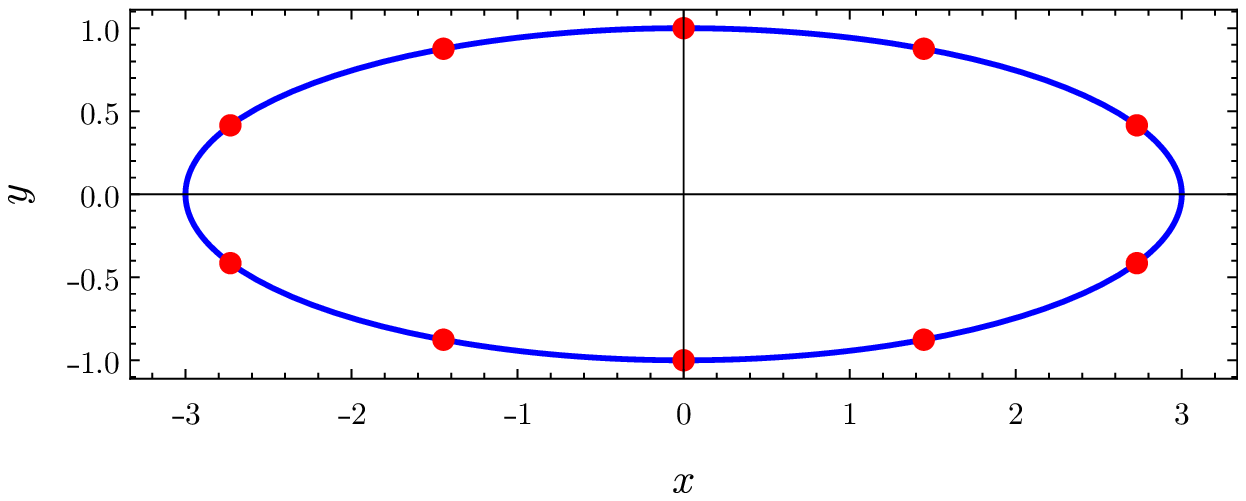} \\
\caption{Equilibrium configurations of $10$ charges on an ellipse with $b=3$ for $s=1$. 
The configurations are displayed according to increasing energy, from left to right and 
from top to down. The configurations in the first row are stable, the ones in the lower row are unstable.}
\label{fig_ellipse_2}
\end{center}
\end{figure}

\begin{figure}[!htb]
\begin{center}
\includegraphics[width=5cm]{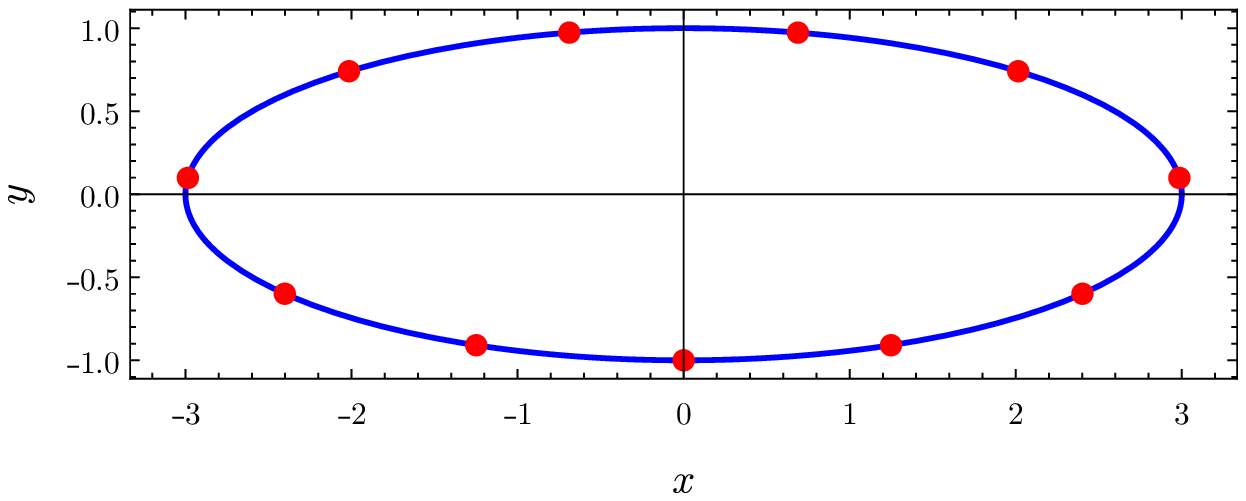} \hspace{1cm} 
\includegraphics[width=5cm]{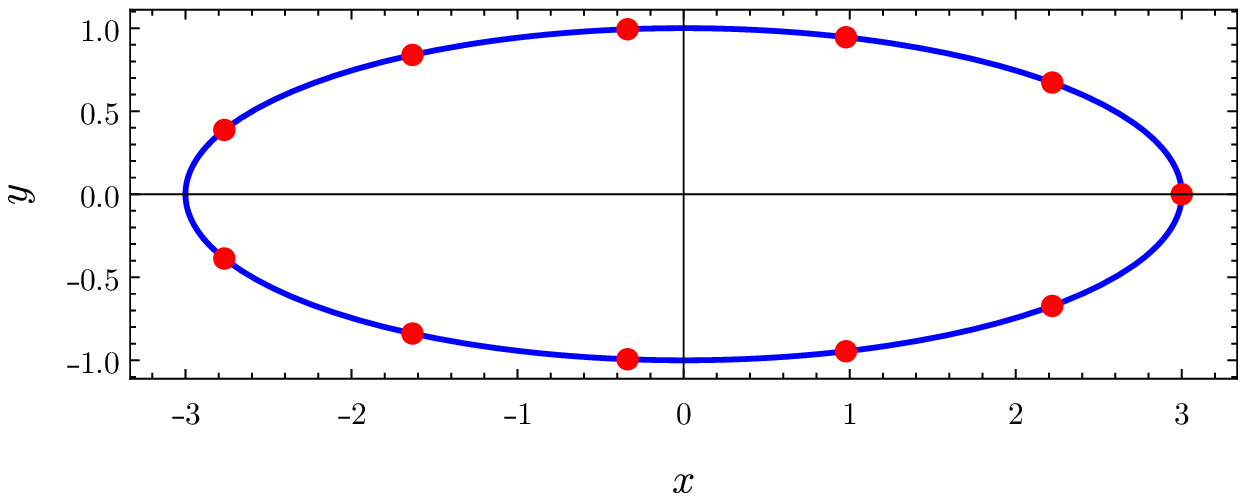}
\caption{Equilibrium configurations of $11$ charges on an ellipse with $b=3$ for $s=1$. The left configuration is stable, the right one is unstable.}
\label{fig_ellipse_3}
\end{center}
\end{figure}

An example of these configurations is displayed in Figs. \ref{fig_ellipse_2} and \ref{fig_ellipse_3}  for the case $N=10$ and $N=11$, 
obtained numerically for an ellipse with $b=3$, for $s=1$. In general, for a given $N$, there are at least two configurations: 
if $N$ is even, the configuration of minimal energy (global minimum) is symmetric with respect to reflections about the 
two principal axes and with respect to the origin, whereas a second (either stable or unstable) configuration of slightly 
higher energy is symmetric only with respect to reflections about the vertical axis;  if $N$ is odd, the configuration 
of minimal energy is even with respect to reflections about the vertical axis, whereas a second (unstable) configuration 
is symmetric with respect to reflections about the horizontal axis.

The different configurations for a give number of charges have been obtained running the algorithm a few times, 
for the same number of charges, starting from randomly generated configurations.

We have found out that the configurations with lowest energy are separated by a gap that decays exponentially with $N$: 
in general the numerical results are fitted 
\begin{eqnarray}
\left. \Delta E\right|_{fit} &=& a e^{-\beta N - \gamma / N - \delta / N^2} 
\end{eqnarray}

\begin{figure}[t]
\begin{center}
\includegraphics[width=7cm]{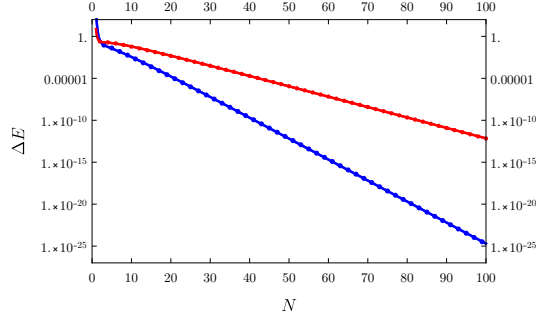}
\caption{Energy gap between the two lowest configurations for an ellipse with $b=2$ and charges interacting via a Coulomb potential ($s=1$), as a function of the number
of charges on the ellipse. The lower (higher) points correspond to configurations with $N$ odd (even). The solid lines are  fits.}
\label{fig_ellipse_4}
\end{center}
\end{figure}

For the case in Fig.~\ref{fig_ellipse_4}, corresponding to $b=2$ and $s=1$, the fits (solid lines in the plot) are
\begin{eqnarray}
\Delta E_{odd}(N)  &\approx& 2.935 \  e^{-0.578 N-\frac{11.148}{N}+\frac{18.292}{N^2}} \\
\Delta E_{even}(N) &\approx& 2.224 \  e^{-0.287 N-\frac{7.879}{N}+\frac{9.392}{N^2}} 
\end{eqnarray}

\begin{figure}[t]
\begin{center}
\includegraphics[width=6cm]{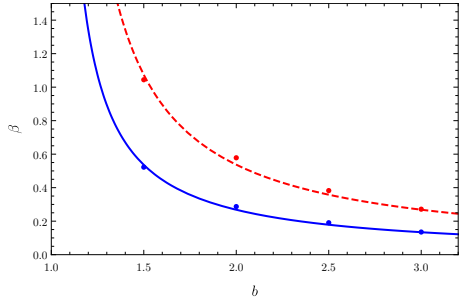}
\caption{$\beta$ as a function of the major semiaxis, for $s=1$. The lines are the fits for the even (solid) and odd (dashed) configurations.}
\label{fig_ellipse_5}
\end{center}
\end{figure}

In Fig.~\ref{fig_ellipse_5} we plot the values of $\beta$ obtained from the fits above, for the cases $s=1$ and $b=1.5,2,2.5,3$, 
both for configurations with even (blue dots) or odd (red dots) number of charges. The lines (solid and dashed) correspond to the simple fit
\begin{eqnarray}
\beta(s,b) &=& \frac{\bar{\beta}(s)}{b-1}
\end{eqnarray}

In particular, for the case in the figure, we have
\begin{eqnarray}
\bar{\beta}_{even}(1) &=& 0.268 \\
\bar{\beta}_{odd}(1) &=& 0.537
\end{eqnarray}

We observe a mild dependence on $s$ of $\bar{\beta}(s)$ since
\begin{eqnarray}
\bar{\beta}_{even}(2) &=& 0.254 \\
\bar{\beta}_{odd}(2) &=& 0.513 
\end{eqnarray}
and
\begin{eqnarray}
\bar{\beta}_{even}(3) &=& 0.250 \\
\bar{\beta}_{odd}(3) &=& 0.509
\end{eqnarray}

\begin{figure}[!htb]
\begin{center}
\includegraphics[width=5.8cm]{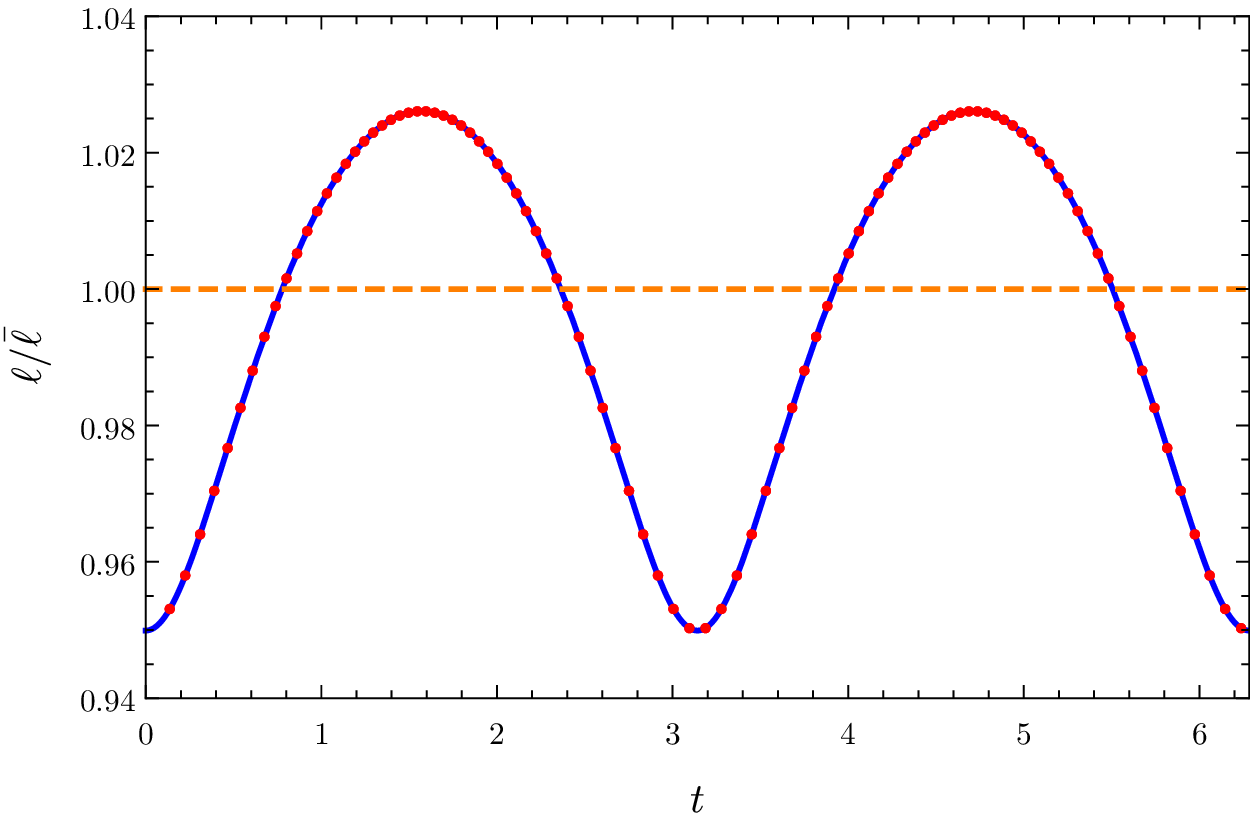} 
\hspace{1cm}
\includegraphics[width=6cm]{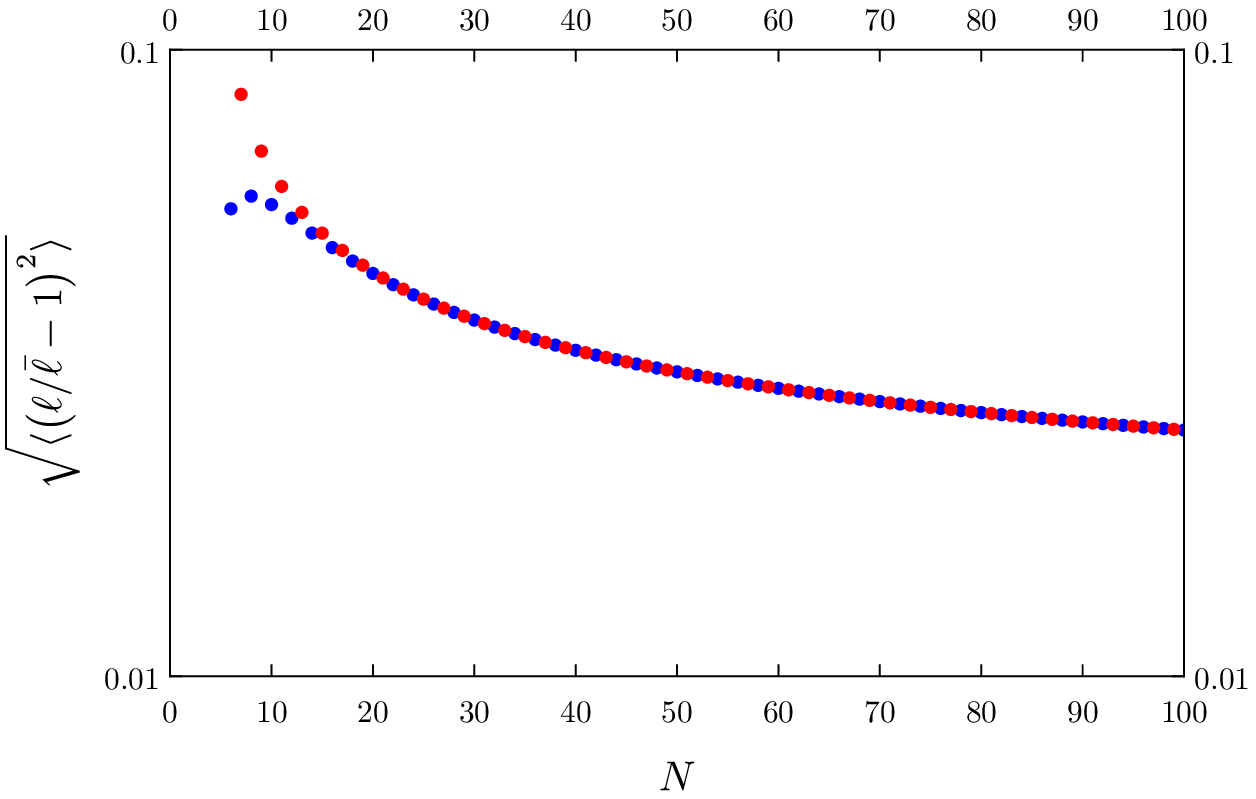}
\caption{Left plot: $\ell /\bar{\ell}$ for the configuration of $100$ charges distributed on the ellipse with major semiaxis $b=2$ and with $s=1$. The points correspond to the distance between a charge and its neighbor, divided by the average length $\ell/N$; the curve is a fit with trigonometric functions.
Right plot: Standard deviation of $\ell /\bar{\ell}$ as a function of the number of charges distributed 
on the ellipse with major semiaxis $b=2$ and with $s=1$. The blue (red) points correspond even (odd) values of $N$. 
}
\label{fig_ellipse_6}
\end{center}
\end{figure}

The numerical results that we have obtained allow us to address the question of how the charges tends to distribute over the ellipse 
as $N$ grows. In the left plot of Fig.~\ref{fig_ellipse_6} we plot the distance between consecutive charges 
normalized by the average distance $\bar{\ell} = \Gamma/N$ ($\Gamma$ is the perimeter of the ellipse), for $N=100$ charges and for $b=2$ and $s=1$ 
(note that $\langle \ell/\bar{\ell} \rangle =  1$).

In the right plot of Fig.~\ref{fig_ellipse_6} we display the standard deviation 
$\Delta_\ell \equiv \sqrt{\langle \left(\ell/\bar{\ell}-1\right)^2 \rangle}$ as a function of the number of 
charges, up to $100$ charges. The slow decay of $\Delta_\ell$ for $N \gg 1$ signals the tendency of the charges 
to distribute uniformly on the ellipse.

\begin{figure}[!htb]
\begin{center}
\includegraphics[width=5.8cm]{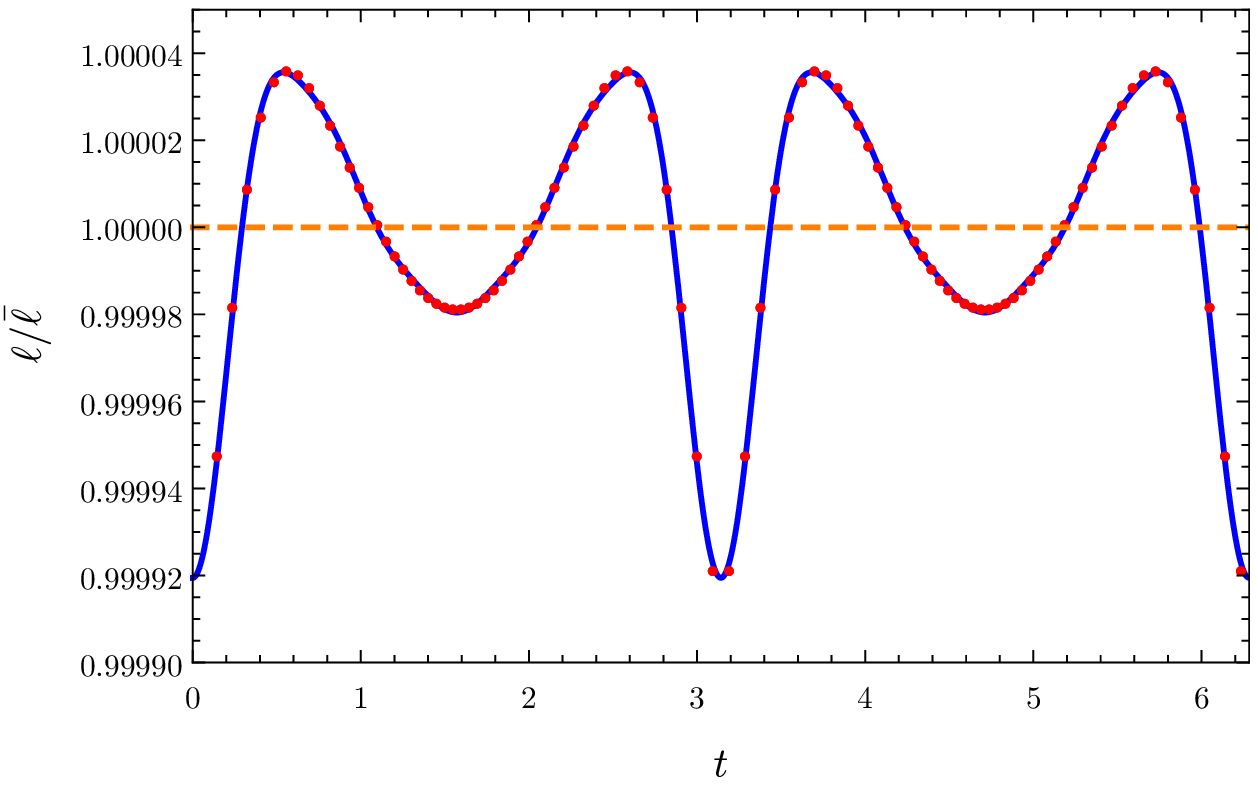} 
\hspace{1cm}
\includegraphics[width=6cm]{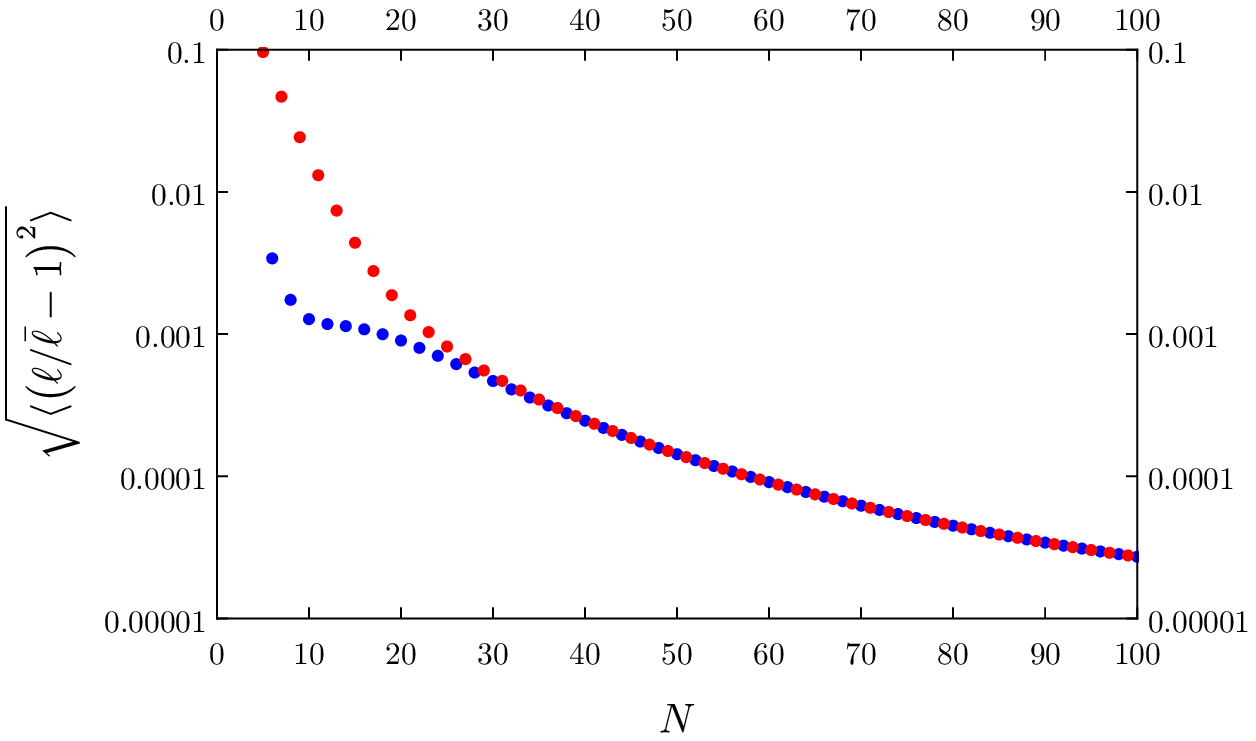}
\caption{Left plot: $\ell /\bar{\ell}$ for the configuration of $100$ charges distributed on the ellipse with major semiaxis $b=2$ and with $s=2$. The points correspond to the distance between a charge and its neighbor, divided by the average length $\ell/N$; the curve is a fit with trigonometric functions.
Right plot: Standard deviation of $\ell /\bar{\ell}$ as a function of the number of charges distributed on the ellipse with major semiaxis $b=2$ and with $s=2$. The blue (red) points correspond even (odd) values of $N$. 
}
\label{fig_ellipse_7}
\end{center}
\end{figure}

The analogous of Fig.~\ref{fig_ellipse_6}, for the case of $s=2$, are shown in Fig.~\ref{fig_ellipse_7}. 
In the left plot, corresponding to $N=100$, we see that the deviations from uniformity are very small ($\max(|\ell/\bar{\ell}-1|) \approx 10^{-4} - 10^{-5}$); moreover, the system
approaches the uniform distribution rather rapidly, as one can see from the right plot, representing the standard deviation $\Delta \ell$.

In addition to the previous results, we have also used  the numerical results obtained for the different cases to estimate with good accuracy ($3-4$ decimals) 
the leading behavior of the electrostatic energy for $N \gg 1$, obtained in  \cite{Martinez04}.

\subsection{Straight needle}
\label{sub:needle}

The problem of how electric charges distribute over a  finite straight wire has been first discussed by Maxwell \cite{Maxwell}, 
and (much) later rediscovered by several authors (see Refs.~\cite{Griffiths96,Griffiths01b,Jackson00,Jackson02,Andrews97}). 
In particular Griffiths and collaborators ~\cite{Griffiths96,Griffiths01b} have studied the discrete version of this problem, 
which they refer to as "bead model", in which $N$ equal charges distribute on a finite straight wire of size $2a$, under the action
of the repulsive Coulomb force between them. These authors found that the charges tend slowly towards a uniform distribution as $N$ grows, 
with a mild logarithmic behavior of the density at the ends of the segment.

Our first approach to the problem is considering a uniform distribution of $N$ charges on the needle of unit length
($|x|\leq 1/2$).  In this case the charges are located at $x_i = -1/2+ (i-1)/(N-1)$, with $i=1, \dots, N$. 

The electrostatic energy for this configuration is
\begin{eqnarray}
E_s(N) = 2 (N-1)^s \sum_{i=2}^N \sum_{j=1}^{i-1} \frac{1}{|i-j|^s} \hspace{0.5cm} , \hspace{0.5cm} s \geq 1
\end{eqnarray}

In particular, for $s=1$, we have
\begin{eqnarray}
E_1(N) = 2 (N-1) N \left(H_N-1\right)
\end{eqnarray}
where $\gamma \approx 0.5772$ is the Euler-Mascheroni constant and $H_N \equiv \sum_{k=1}^N \frac{1}{k}$ is the $N^{th}$ harmonic number.
For $N \rightarrow \infty$, the asymptotic behavior of the energy is
\begin{eqnarray}
E_1(N) \approx 2 N^2 \log N + 2 (\gamma -1) N^2 -2 N \log N + (3-2 \gamma ) N+  \dots
\end{eqnarray}
where the leading term agrees with the exact expression of refs.~\cite{Martinez04, Borodachov12}.

Similarly, for $s = 2$ and for $N \rightarrow \infty$ one has
\begin{eqnarray}
E_2(N) &\approx& \frac{\pi ^2 N^3}{3}-2 N^2 \log N -\frac{2}{3} \left(3+3 \gamma +\pi ^2\right) N^2 \nonumber \\
&+& 4 N \log N  + N \left(\frac{\pi ^2}{3}+4 \gamma +4\right) + \dots
\end{eqnarray}

Once again the leading term in $E_2(N)$ for $N \rightarrow \infty$ coincides with the general expression of refs.~\cite{Martinez04, Borodachov12}.

\begin{figure}[!htb]
\begin{center}
\includegraphics[width=5.8cm]{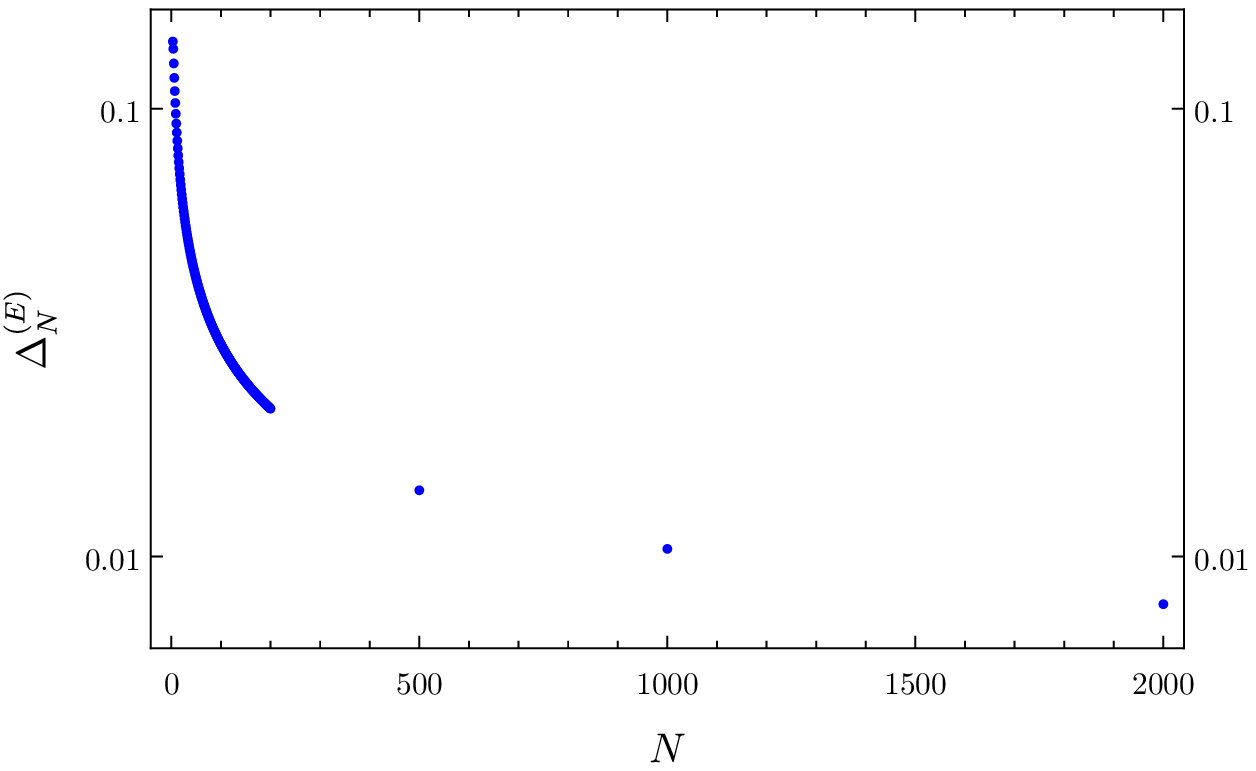} 
\hspace{1cm}
\includegraphics[width=6cm]{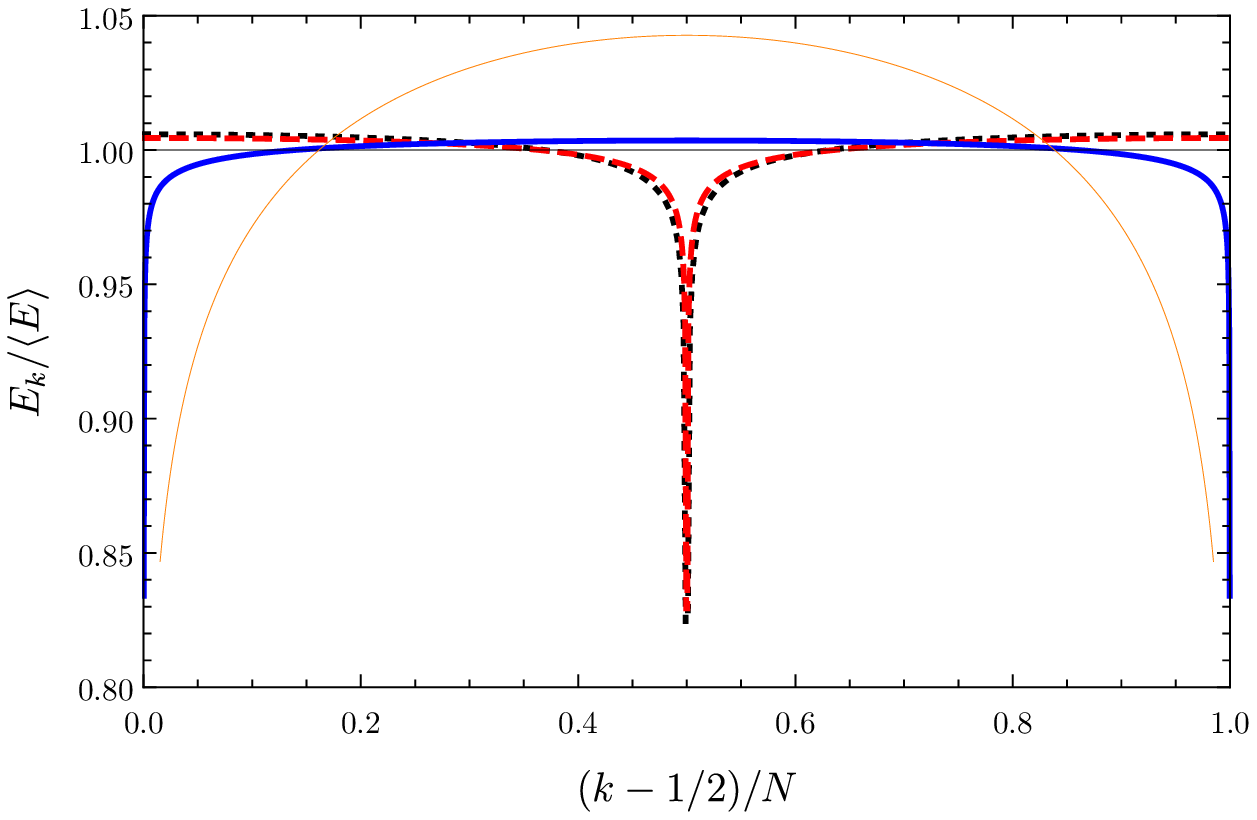}
\caption{Left plot: standard deviation for the electrostatic energy of a system of $N$ charges on a unit straight needle as a function of the number of charges for $s=1$;
Right plot: electrostatic energy of the charges on the needle for systems with $N=500,1000,2000$ charges (black dotted, red dashed and solid blue 
lines respectively, for $s=1$. The thin orange line is the energy corresponding to the uniform distribution of $2000$ charges.
}
\label{fig_needle_1}
\end{center}
\end{figure}

We have obtained the optimal configurations of systems with different number of charges ($N=2,3, \dots, 200$ and $N=500,1000,2000$) on the unit straight needle, using the Newton method.
In the left plot of Fig.~\ref{fig_needle_1} we report the standard deviation of the Coulomb electrostatic energy of a system of $N$ charges, $\Delta_N^{(E)}$, as a function of $N$;  one can appreciate that $\Delta_N^{(E)}$ decreases monotonically with $N$, suggesting that it may tend to $0$ as $N \rightarrow \infty$.

In the right plot of Fig.~\ref{fig_needle_1} we report the behavior of the electrostatic energy of a single charge, normalized with the average electrostatic energy of the system, for 
configurations with large number of charges, $N=500,1000, 2000$. In this case we see that, going from $1000$ to $2000$ charges, the electrostatic energy undergoes a sudden change: while for 
$N \lesssim 1000$ the charges at the center of the needle have a lower energy, the situation dramatically changes for $N \gtrsim 1000$, with the charges at the ends of the needle now being
at lower energy (the thin horizontal line at $E_k/\langle E \rangle = 1$ is just meant to guide the eye, whereas the thin orange line corresponds to the electrostatic energy of a perfectly
uniform distribution).

\begin{figure}[!htb]
\begin{center}
\includegraphics[width=5.8cm]{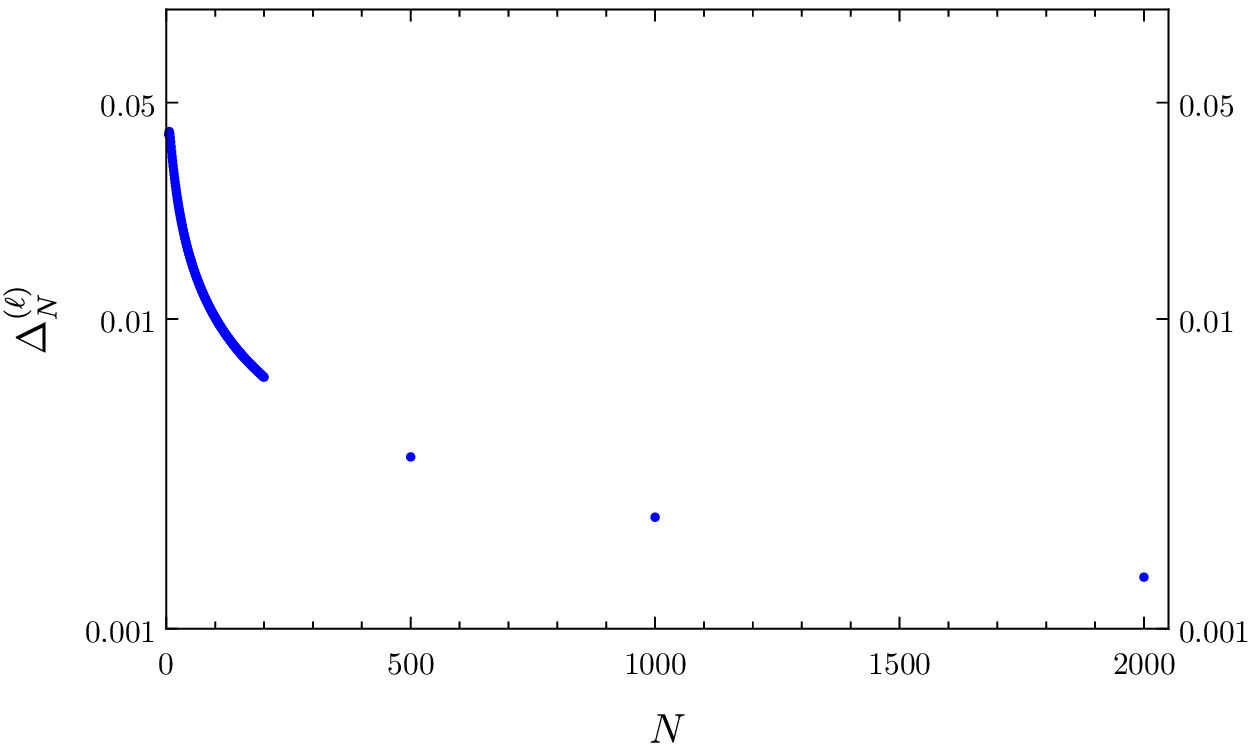} 
\hspace{1cm}
\includegraphics[width=6cm]{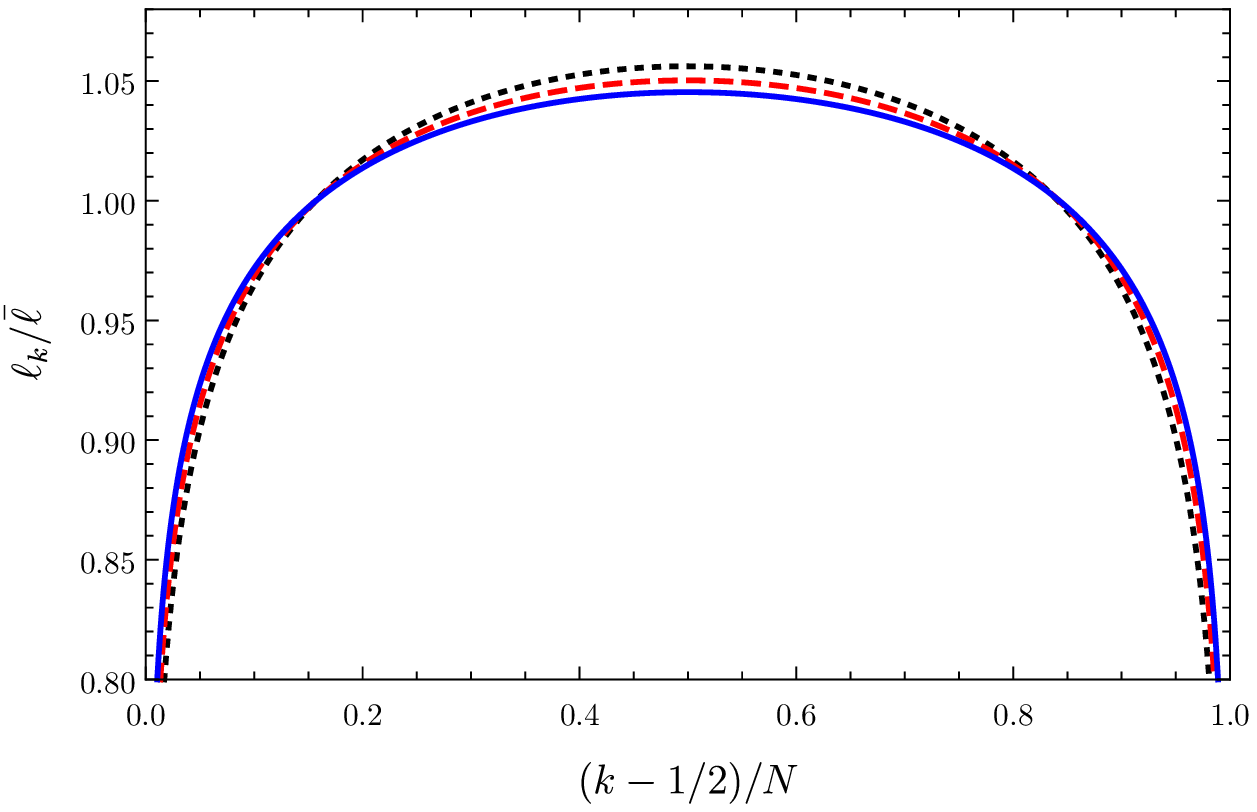}
\caption{Left plot: standard deviation for the interparticle distance of a system of $N$ charges as a function of the number of charges for $s=1$;
Right plot: normalized interparticle distance of the charges on the needle for systems with $N=500,1000,2000$ charges (black dotted, red dashed and solid blue lines respectively, for $s=1$. 
}
\label{fig_needle_2}
\end{center}
\end{figure}

A similar analysis is performed in Fig.~\ref{fig_needle_2} for the standard deviation of the interparticle distance of a system of 
$N$ charges on the unit needle, for $s=1$; as we appreciate from the left plot, also in this case the standard deviation decreases 
monotonically with $N$ and seems to tend to $0$ for $N \rightarrow \infty$. This clearly suggests that the system is approach a 
uniform distribution for $N \rightarrow \infty$. In the right plot of Fig.~\ref{fig_needle_2} we compare the distributions of (normalized) 
interparticle distances for configurations corresponding to $N= 500, 1000, 2000$. Unlike the case of the electrostatic energy, we 
see that the three curves are qualitatively similar and correspond to configurations where the charge density at the center of the 
needle is lower.


\subsection{Cardioid}
\label{sub:cardiod}

Our last example is the cardioid defined by the equation
\begin{eqnarray}
\vec{R}(t) = \sqrt{\frac{2}{3}} \ (\cos t + \frac{1}{2} \cos 2t  , \sin t + \frac{1}{2} \sin 2t )  \hspace{.5cm} , \hspace{.5cm}  0 \leq t < 2 \pi
\end{eqnarray}

The reason for studying this curve is to assess the effect, if any, of the cusp on the charge distribution, for $N \rightarrow \infty$.
We have obtained numerically all the equilibrium configurations for charges ranging from $N=2$ to $N=200$, for potentials
corresponding to $s=1,2,3$. The configurations with $500$ and $1000$ charges have also been  calculated for the cases $s=1,2$.

\begin{figure}[t]
\begin{center}
\includegraphics[width=4cm]{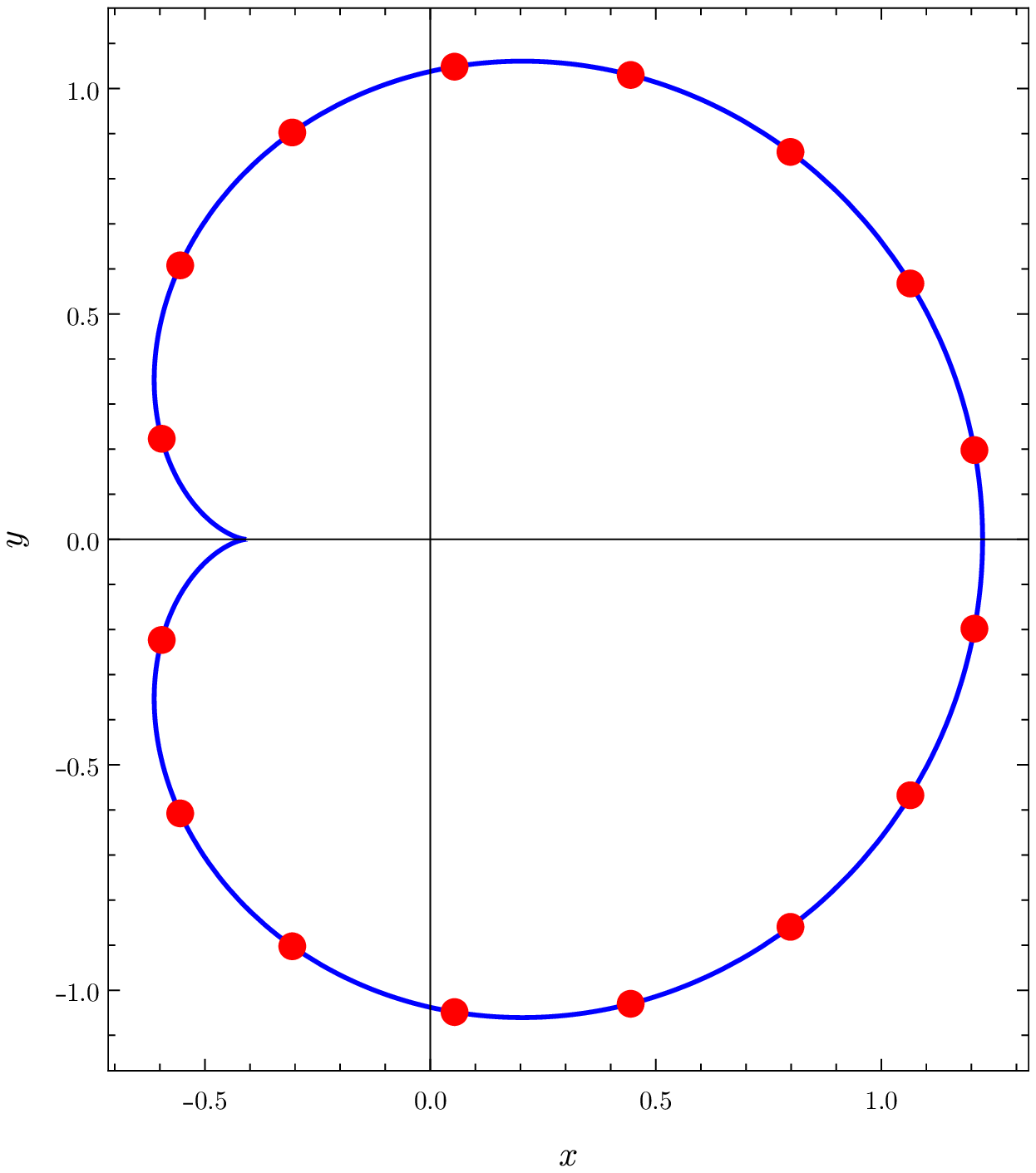} \hspace{1cm} 
\includegraphics[width=4cm]{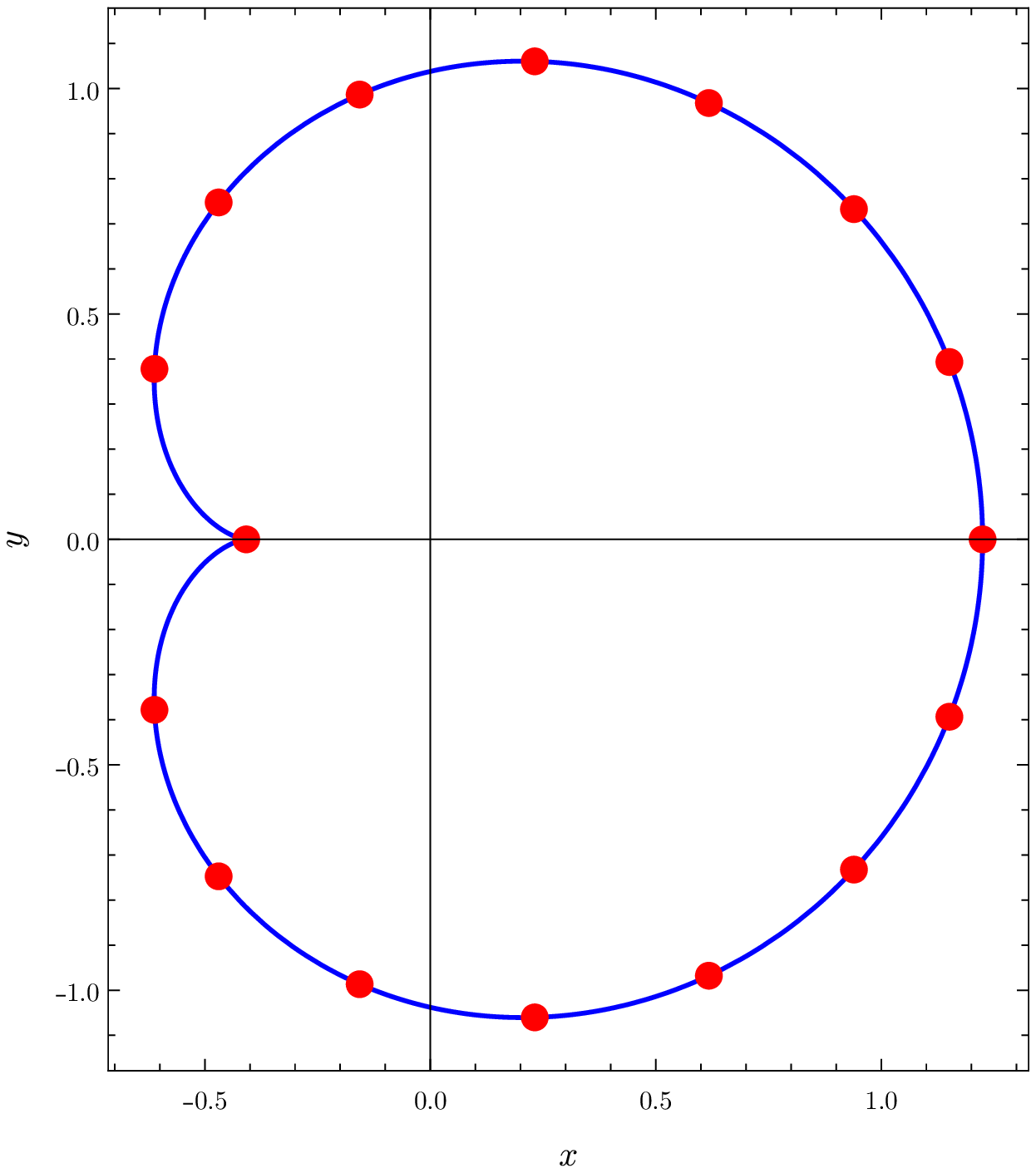} \\
\caption{Equilibrium configurations of $16$ charges on the cardioid for $s=1$. 
The configuration in the left plot has a lower energy.}
\label{fig_Cardioid_1}
\end{center}
\end{figure}

\begin{figure}[t]
\begin{center}
\includegraphics[width=4cm]{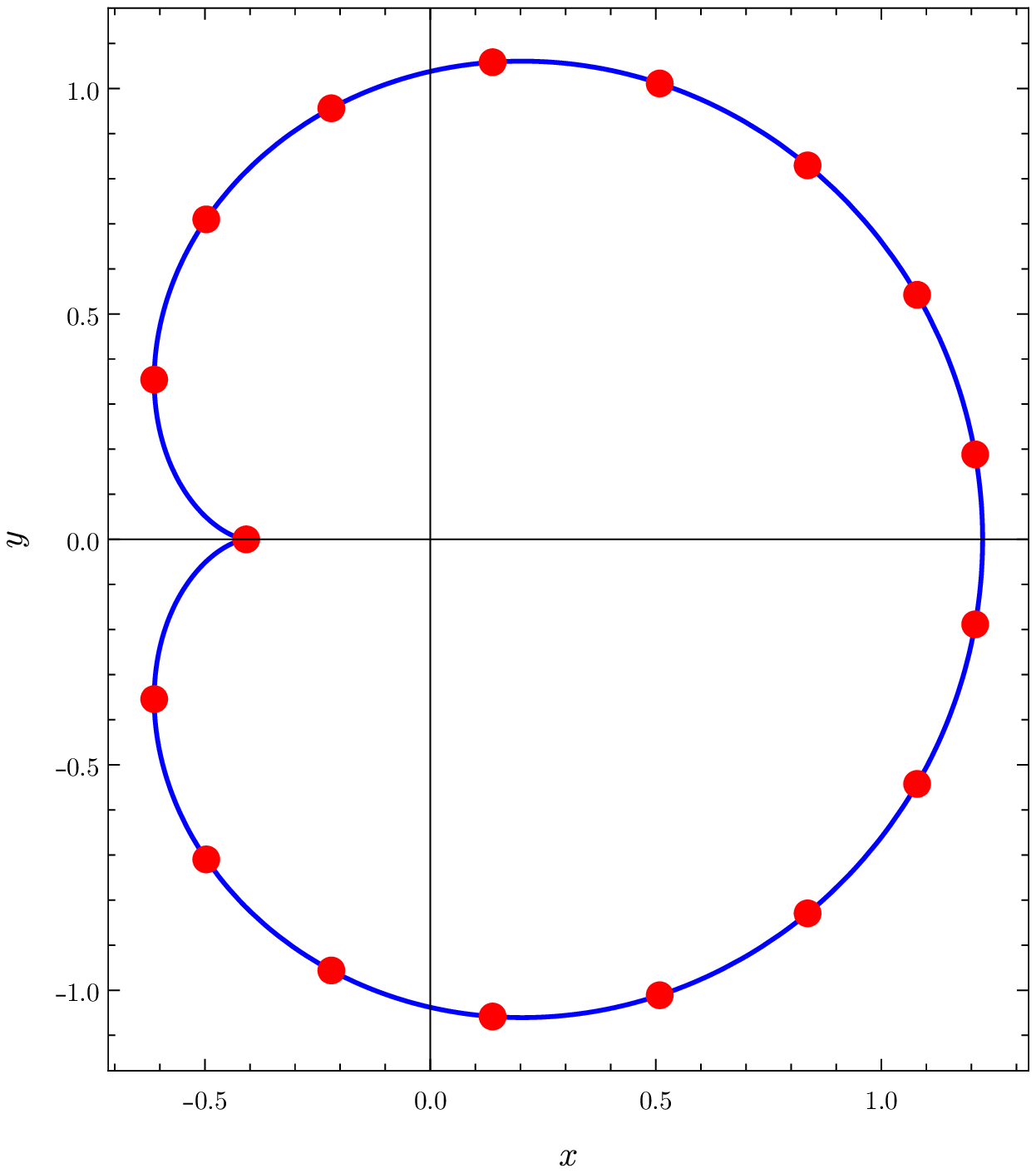} \hspace{1cm} 
\includegraphics[width=4cm]{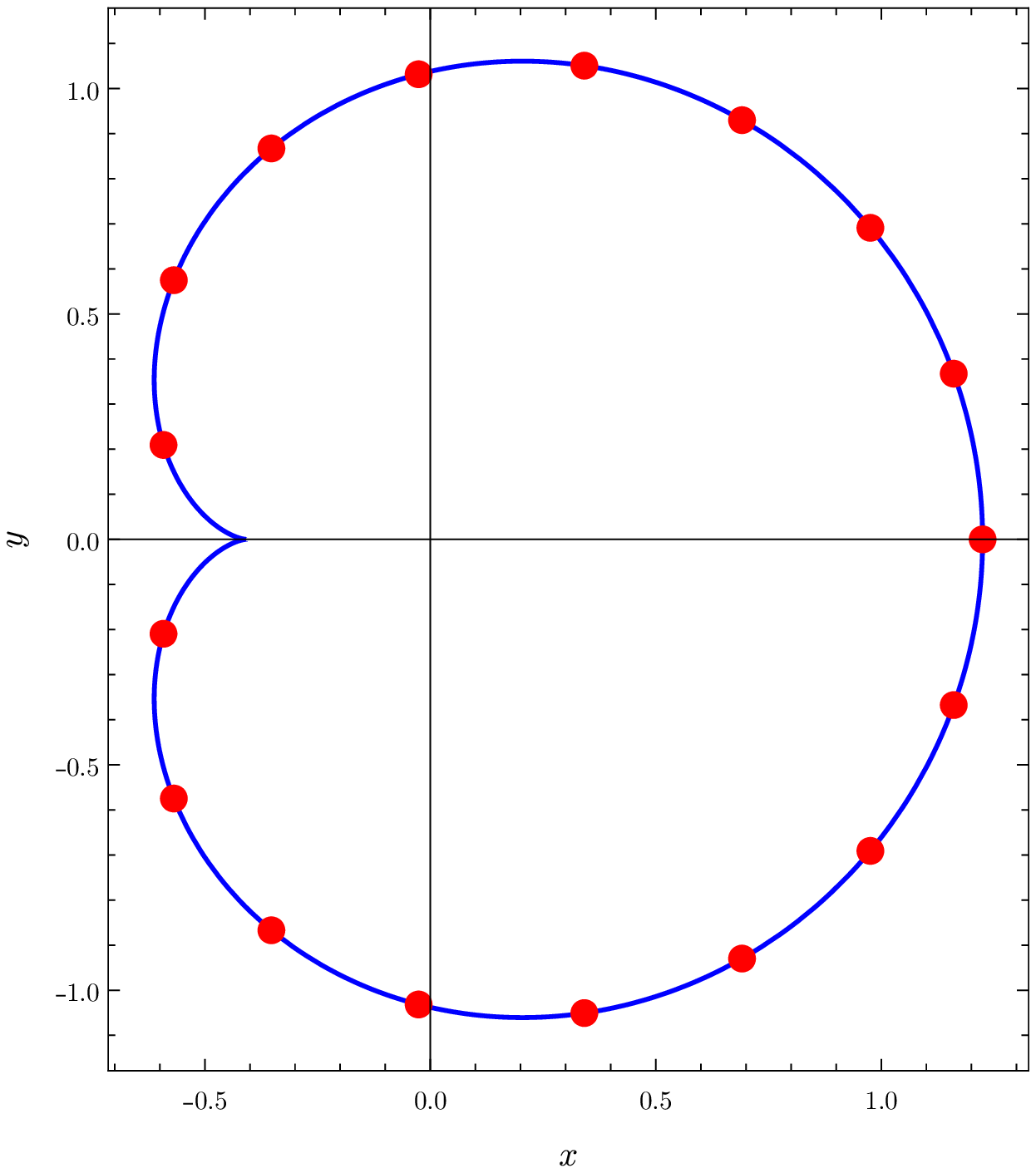} \\
\caption{Equilibrium configurations of $17$ charges on the cardioid for $s=1$. 
The configuration in the left plot has a lower energy.}
\label{fig_Cardioid_2}
\end{center}
\end{figure}

\begin{figure}[t]
\begin{center}
\includegraphics[width=7cm]{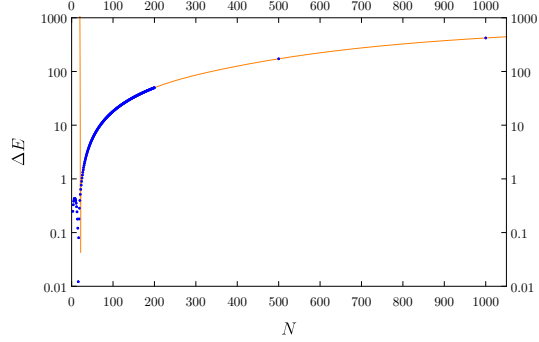}
\caption{Energy gap between the two lowest configurations for the cardioid with charges interacting via a Coulomb potential ($s=1$), 
as a function of the number of charges on the ellipse. The solid line is a fit.  }
\label{fig_Cardioid_3}
\end{center}
\end{figure}

For all the cases studied we have found out that there are always two different equilibrium configurations:
one in which the cusp is not occupied by any charge and one in which a charge sits on the cusp itself. 
The numerical results show that the first configuration has lower energy only at low density (for $s=1$, $N \geq 16$). 
In Figs.~\ref{fig_Cardioid_1} and \ref{fig_Cardioid_2} we show the configurations for $s=1$ with $16$ and $17$ charges respectively,
where the left plot in each figure corresponds to the configuration of lower energy.
Unlike for the case of the ellipse, where the energy gap between the lowest configurations was decreasing
exponentially with the number of charges, in this case the gap grows for $N \gg 1$ (see Fig.~\ref{fig_Cardioid_3}).

\begin{figure}[t]
\begin{center}
\includegraphics[width=6cm]{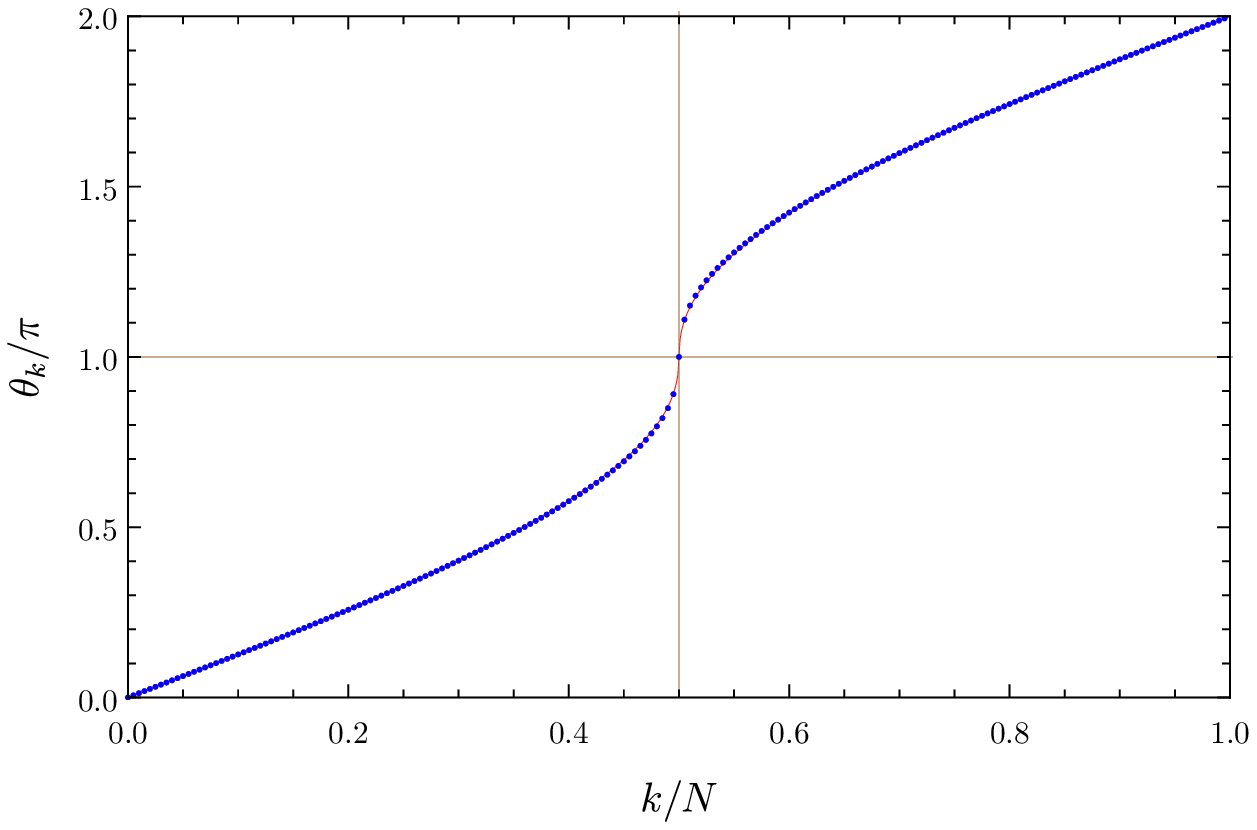}  \hspace{1cm}
\includegraphics[width=6cm]{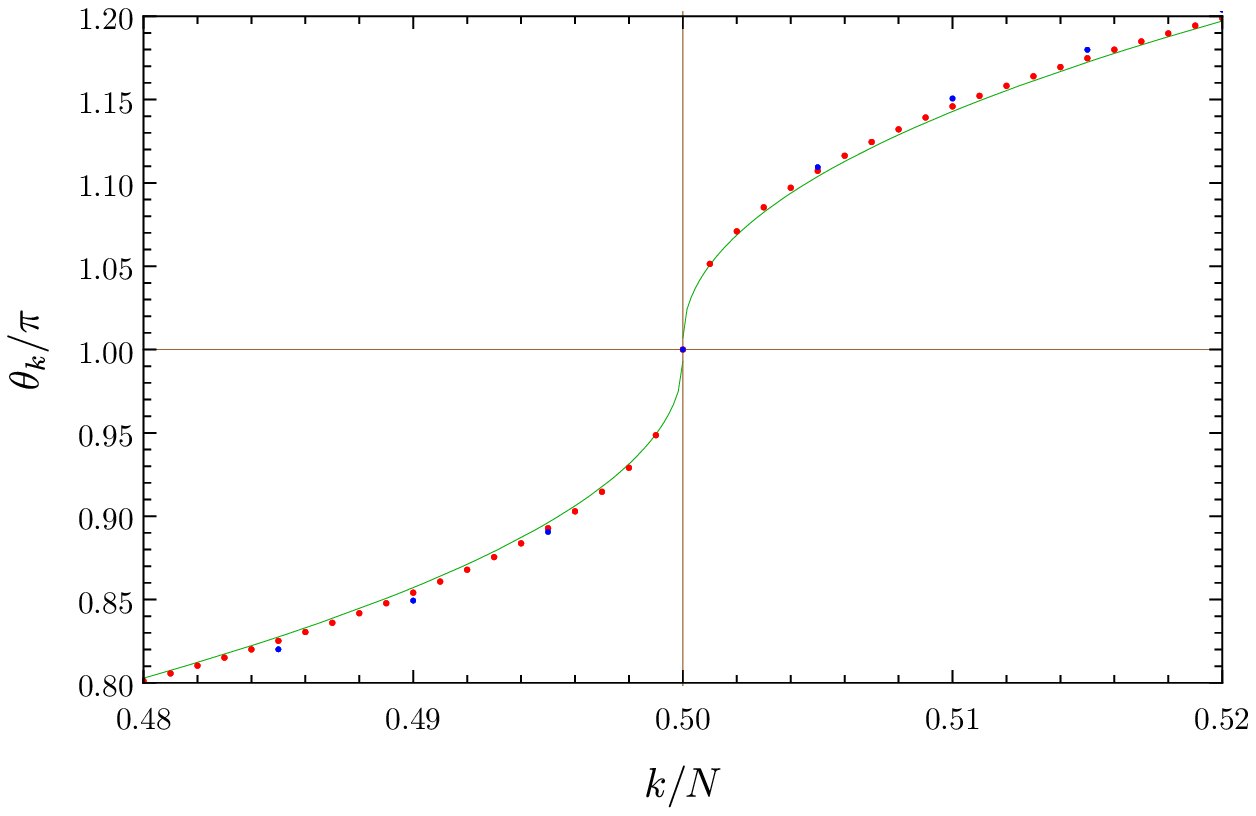}  
\caption{Left plot: Polar angles for the lowest energy configurations of $500$ and $1000$ charges (dots and solid line respectively) with $s=1$, as a function of the $k/N$ ($k$ is the integer index labelling each charge). Right plot: detail of the left plot (the red dots now correspond to the configuration of $1000$ charges, whereas the solid line is the two paramer fit (\ref{fitthetacardioid}))}
\label{fig_Cardioid_4}
\end{center}
\end{figure}

An interesting behavior is found when we plot the polar angles, ordered in a monotonic sequence, 
for the lowest energy configurations of $500$ and $1000$ charges 
as a function of $k/N$, where $k$ is an integer number that labels each charge ($k=1,\dots N$). 
The left plot of Fig.~\ref{fig_Cardioid_4}, corresponding to charges interacting via the Coulomb potential, 
suggests that two sets, respectively represented by dots and by a solid line, have the same behavior.

A simple two parameters fit of the points corresponding to the configurations of $1000$ charges
\begin{eqnarray}
\Theta(x) = \left\{ \begin{array}{ccc}
f(x)     & , &   0 \leq x \leq 1/2  \\
2-f(1-x) & , &  1/2 < x \leq 1  \\
\end{array}
\right.
\end{eqnarray}
where
\begin{eqnarray}
f(x) = (1-\sqrt{1-2 x}) \left[ a + b (x-1/2) \right] \ \ \ , \ \ \ 0 \leq x \leq \frac{1}{2}
\label{fitthetacardioid}
\end{eqnarray}
with 
\begin{eqnarray}
a = 0.99315  \ \ \ ,  \ \ \ b = -0.51928
\end{eqnarray}
describes particularly well the observed behavior. 
The right plot of Fig.~\ref{fig_Cardioid_4} shows a detail of the left plot, around the region of the cusp, where the
solid line corresponds to the fit (\ref{fitthetacardioid}) and the red dots correspond to the angles for the configuration
with $1000$ charges.

\begin{figure}[t]
\begin{center}
\includegraphics[width=5.8cm]{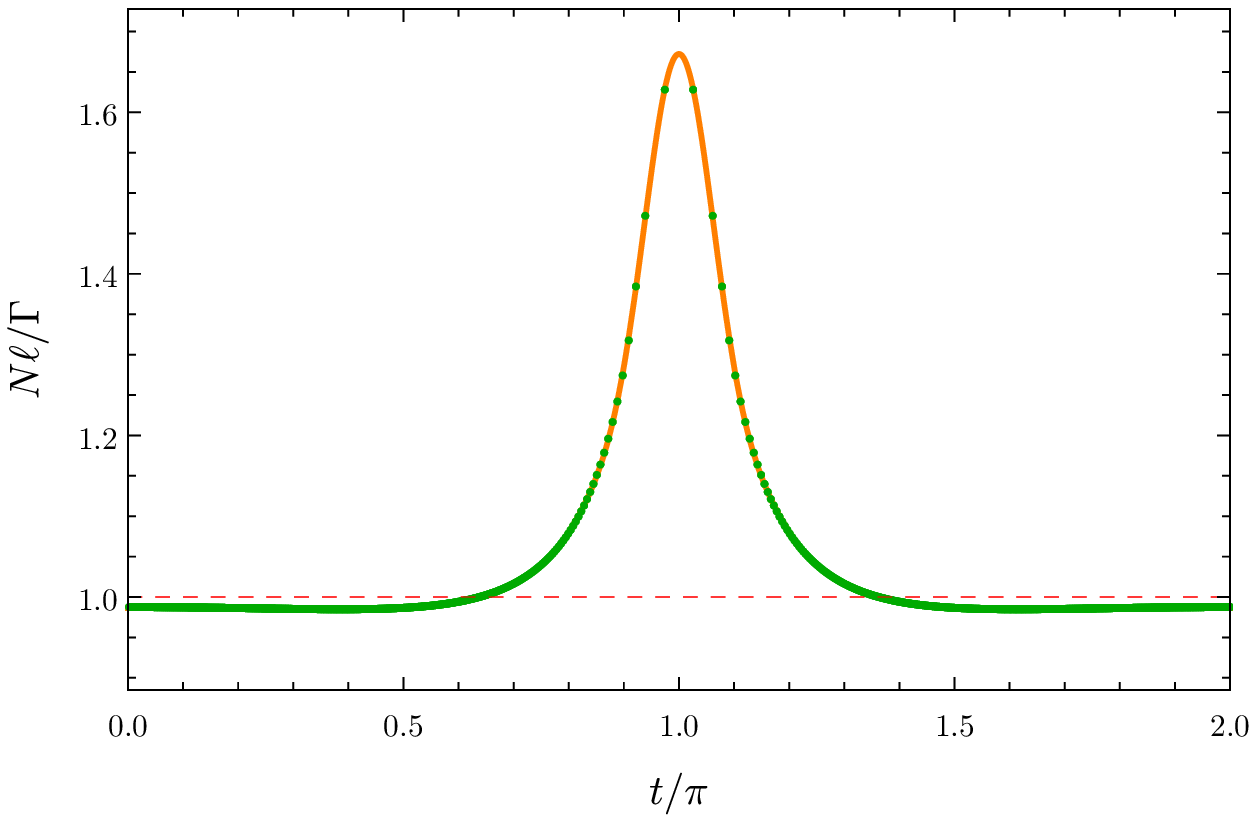} 
\hspace{1cm}
\includegraphics[width=6cm]{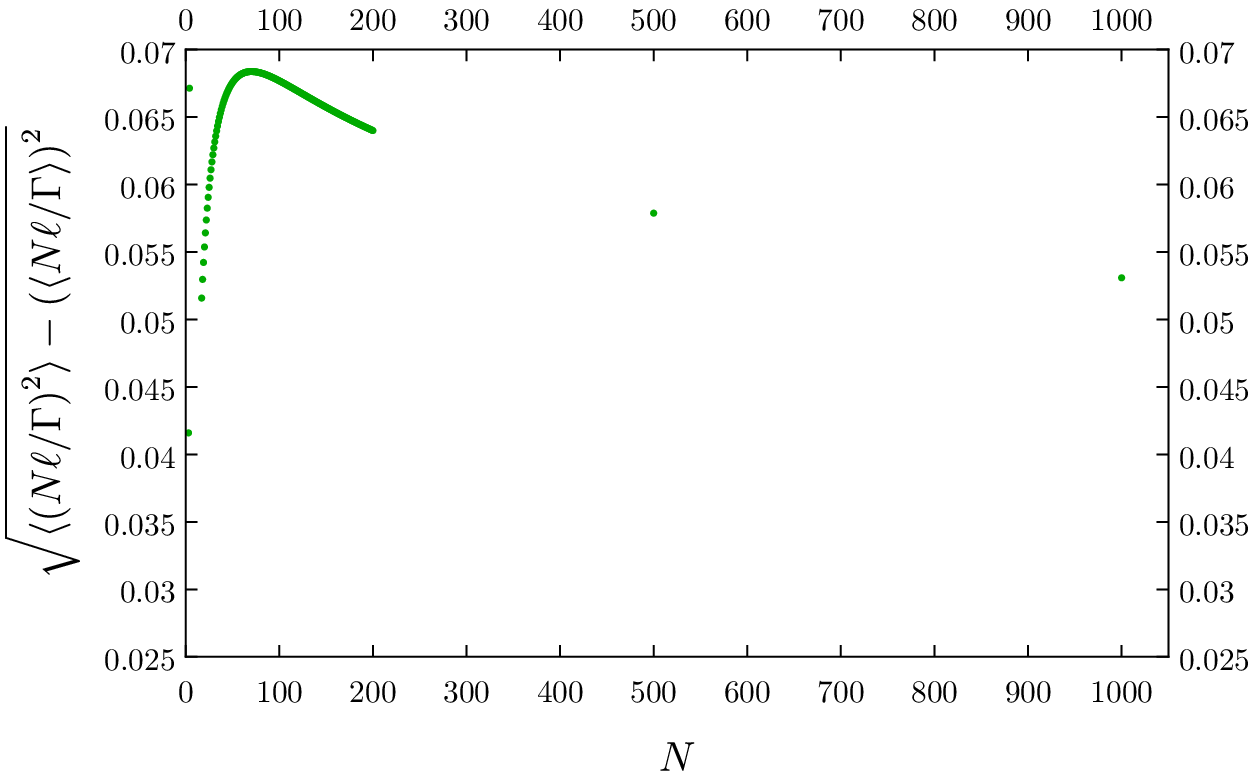}
\caption{Left plot: $\ell /\bar{\ell}$ for the configuration of $1000$ charges distributed on the cardioid and interacting with the 
Coulomb potential ($s=1$). The points correspond to the distance between a charge and its neighbor, divided by the average length 
$\ell/N$; the solid curve is a gaussian fit.
Right plot: Standard deviation of $N \ell /\Gamma$ as a function of the number of charges distributed on the cardioid with $s=1$.
}
\label{fig_Cardioid_5}
\end{center}
\end{figure}

\begin{figure}[t]
\begin{center}
\includegraphics[width=5.8cm]{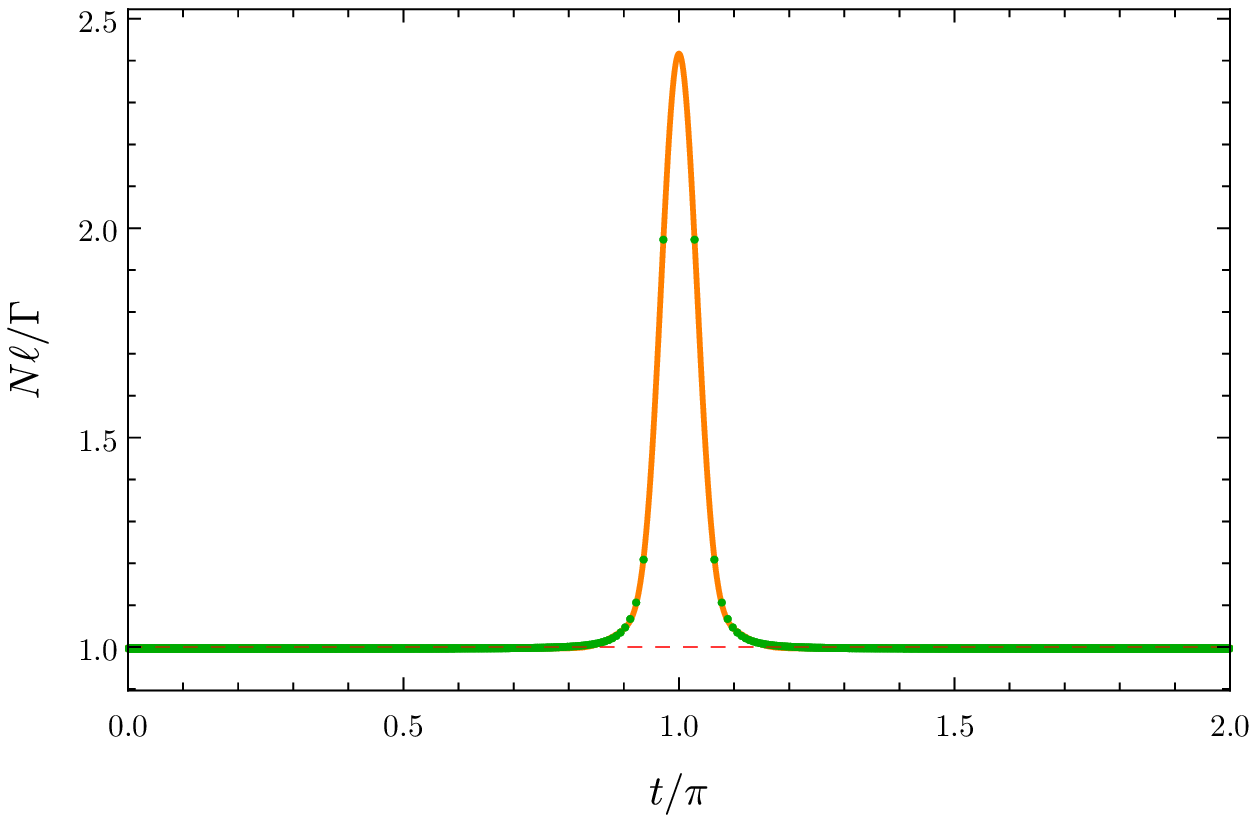} 
\hspace{1cm}
\includegraphics[width=6cm]{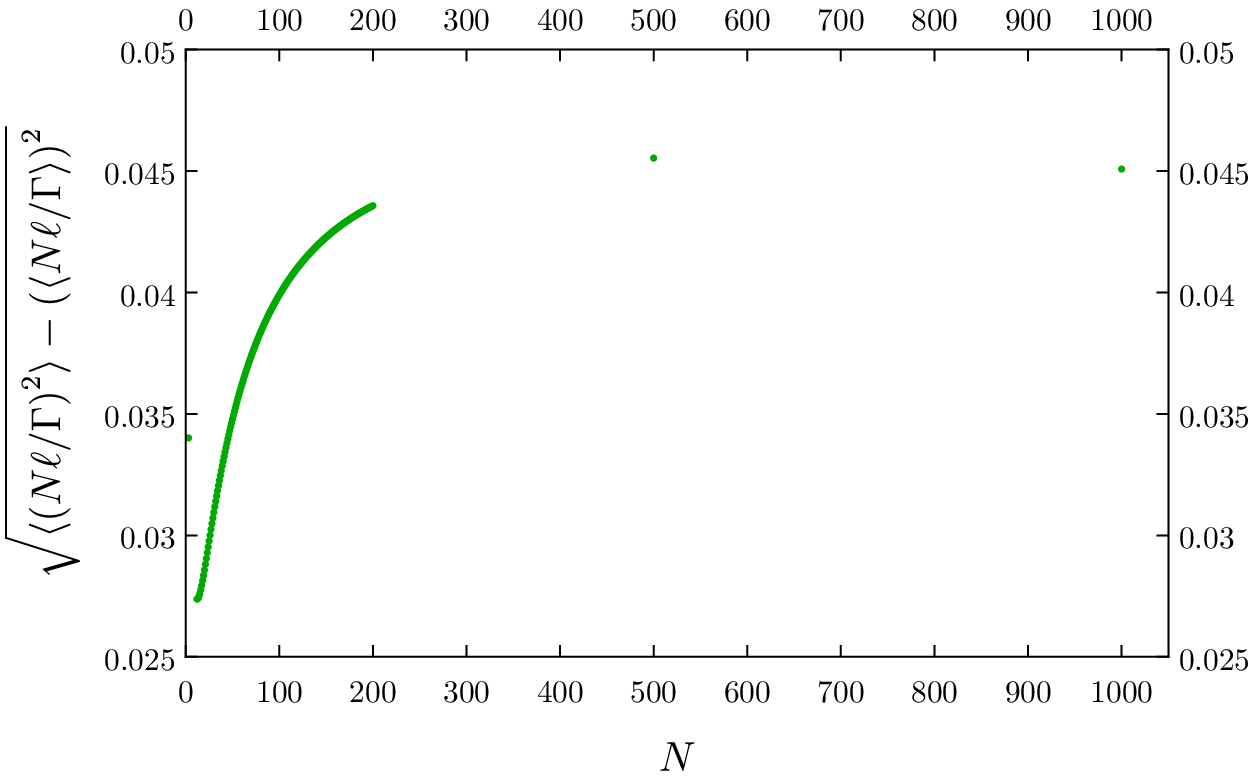}
\caption{Left plot: $\ell /\bar{\ell}$ for the configuration of $1000$ charges distributed on the cardioid and interacting with a potential with $s=2$.  The points correspond to the distance between a charge and its neighbor, divided by the average length 
$\ell/N$; the solid curve is a gaussian fit.
Right plot: Standard deviation of $N \ell /\Gamma$ as a function of the number of charges distributed on the cardioid with $s=2$.
}
\label{fig_Cardioid_6}
\end{center}
\end{figure}

In the left plot of Fig.~\ref{fig_Cardioid_5} we plot the distance between consecutive charges on the cardioid, normalized by the average distance ($\ell/(\Gamma/N)$), for the configuration of $1000$ charges, and $s=1$ (Coulomb potential). In this case one observes that 
in the region of the cusp ($t \approx \pi$) the distance deviates sensibly from the uniform distribution (i.e. from $\ell/(\Gamma/N) \approx 1$).
In the right plot the standard deviation of the interparticle distance is plotted as a function of the number of charges, for $N=2,3, \dots, 200$ and $N=500,1000$. The standard deviation reaches a maximum at $N \approx 100$, followed by a slow decay. 

Fig.~\ref{fig_Cardioid_6} is the analogous of Fig.~\ref{fig_Cardioid_5}, for the case of a potential with $s=2$: in this case we observe that
the deviation from uniformity for the configuration of $1000$ charges is more pronounced than in the case of a Coulomb potential (see left plot
of Fig.~\ref{fig_Cardioid_6}). Similarly we observe that the standard deviation of the interparticle distance as a function of the number of charges grows monotonically in the range of charges that we have considered, with a slope that decreases sensiblly  for $500 < N < 1000$. 

\section{Conclusions}
\label{concl}

In this paper we have studied the problem of how $N$ charges that are repelling each other with an electrostatic force distribute on a curve on the plane; 
this problem is the analogous of the celebrated Thomson problem, regarding $N$ charges on the sphere. Previous work for configurations with finite number of charges
has mainly been focussed on the straight needle, with the first analysis dating back to Maxwell~\cite{Maxwell}; the case corresponding to $N \rightarrow \infty$ has
been studied more recently, in particular in refs.~\cite{Martinez04,Borodachov12}, where the leading asymptotic behavior of the electrostatic energy for closed and  
open curves has been obtained. These results, however, do not include the particular case in which cusps are present and do not tell anything on the behavior at 
finite $N$. We have studied the configurations for $N$ charges on the circle, on ellipses of different eccentricity, on a straight needle and on a cardioid, 
using the Newton method to obtain very precise numerical results, for configurations corresponding to different number of charges and different electrostatic forces.

We may resume our findings with the following points:
\begin{itemize}
\item for the case of the circle, it is possible to obtain the leading asymptotic coefficients for the electrostatic energy, derived in refs.~\cite{Martinez04,Borodachov12},
with high accuracy, performing a suitable extrapolation of the precise numerical results for different $N$;
\item for the cases of ellipses of different eccentricity, we see that, at small $N$, there are several equilibrium configurations, both stable and unstable, and that
the number of configurations tends to increase with $s$ ($s=1$ for the Coulomb potential); moreover the energy gap between the lowest two configurations
decreases exponentially with $N$, as $N \rightarrow \infty$;
\item for the case of a straight needle we observe the tendency to approach a uniform distribution for $N \rightarrow \infty$, as already found by other authors; interestingly
we find that the electrostatic energy of the individual charges undergoes a dramatic change at a large value of $N$, $1000 \lesssim N \lesssim 2000$;
\item for the case of a cardioid,  the presence of the cusp affects  greatly the behavior of the system. For all studied number of charges we have found there are two configurations:
for $N \leq 16$ the lowest energy configuration corresponds to  disposing the charges on the cardioid in a way that the cusp is not occupied, whereas for $N \geq 17$ the lowest energy configuration {\sl always} corresponds to the case where the cusp is occupied. The energy gap between these two configurations {\sl increases} with $N$, in sharp contrast with the case
of the ellipse. Even more interestingly, the cusp is seen to affect greatly the distribution of charges on the curve, with a sizeable effect in proximity of the cusp itself; the 
standard deviation of the normalized interparticle distance in this case does not appear to tend to $0$ for $N \rightarrow \infty$, particularly for the case corresponding to 
$s=2$;
 
\end{itemize}

\section*{Acknowledgements}
The research of P.A. was supported by Sistema Nacional de Investigadores (M\'exico).

\end{document}